\documentclass[12pt,onecolumn,journal]{IEEEtran}

\usepackage{epstopdf}
\usepackage{graphicx}
\usepackage{stmaryrd}
\usepackage{amsmath,amsthm,amssymb}
\usepackage{multirow,url}
\usepackage{color,cite}
\usepackage{algorithm}
\usepackage{algorithmicx}
\usepackage{algpseudocode}
\usepackage{setspace}
\newtheorem{lem}{Lemma}

\newtheorem{thm}{Theorem}
\newtheorem{defn}{Definition}

\ifodd 0
\newcommand{\rev}[1]{{\color{blue}#1}} 
\newcommand{\com}[1]{\textbf{\color{red} (COMMENT: #1)}} 
\newcommand{\comg}[1]{\textbf{\color{green} (COMMENT: #1)}}
\newcommand{\response}[1]{\textbf{\color{magenta} (RESPONSE: #1)}} 
\else
\newcommand{\rev}[1]{#1}
\newcommand{\com}[1]{}
\newcommand{\comg}[1]{}
\newcommand{\response}[1]{}
\fi

\begin{document}

\title{Distributed Spectrum Access with Spatial Reuse}

\author{Xu Chen$^{*}$ and Jianwei Huang$^{*}$\footnote{$*$The authors are with the Network Communications and Economics Lab, Department of Information Engineering, the Chinese University of Hong Kong; Email:\{cx008,jwhuang\}@ie.cuhk.edu.hk; Jianwei Huang is the corresponding author. This work is supported by the General Research Funds (Project Number 412710 and 412511) established under the University Grant Committee of the Hong Kong Special Administrative Region, China.}\vspace{-0.8cm}}

\maketitle
\thispagestyle{empty}

\allowdisplaybreaks

\begin{abstract}
Efficient distributed spectrum sharing mechanism is crucial for improving the
spectrum utilization. The spatial aspect of spectrum sharing, however, is less
understood than many other aspects. In this paper, we generalize a recently proposed spatial
congestion game framework to design efficient distributed spectrum
access mechanisms with spatial reuse. We first propose a spatial channel
selection game to model the distributed channel selection problem
with fixed user locations. We show that the game is
a potential game, and develop a distributed learning mechanism that
converges to a Nash equilibrium only based on users' local observations. We then formulate the joint channel and location selection problem
as a spatial channel selection and mobility game, and show that it
is also a potential game. We next propose a distributed strategic
mobility algorithm, jointly with the distributed learning mechanism,
that can converge to a Nash equilibrium. Numerical results show that the Nash equilibria achieved by the  proposed algorithms have only less than $8\%$ performance loss,
compared with the centralized optimal solutions.
 \end{abstract} 

\section{Introduction}
Dynamic spectrum sharing is envisioned as a promising technique to alleviate
the problem of spectrum under-utilization \cite{key-1}. It enables unlicensed
wireless users (secondary users) to opportunistically access the licensed
channels owned by legacy spectrum holders (primary users), and thus
can significantly improve the spectrum efficiency \cite{key-2}.

A key challenge of dynamic spectrum sharing is how to resolve the resource competition
by selfish secondary users in a decentralized fashion. If multiple
secondary users transmit over the same channel simultaneously, it may lead to severe interference
and reduced data rates for all users. Therefore, it is necessary
to design efficient distributed spectrum sharing mechanism.

The competitions among secondary users for common spectrum resources have often
been studied using noncooperative game theory (e.g., \cite{key-3,key-21,key-4,key-22,key-16}).
Nie and Comaniciu in \cite{key-21} designed a self-enforcing distributed
spectrum access mechanism based on potential games. Niyato and Hossain
in \cite{key-4} studied a price-based spectrum access mechanism for
competitive secondary users. F¨¦legyh¨¢zi \emph{et al.} in \cite{key-22} proposed a
two-tier game framework for medium access control (MAC) mechanism
design. Law \emph{et al.} in \cite{key-16} studied the system performance degradation due to users' selfish behaviors in spectrum
access games.

When not knowing spectrum information such as channel availabilities, secondary users need to learn the environment and adapt the spectrum access decisions accordingly. Han \emph{et al.} in \cite{key-25} and Maskery \emph{et al.} in \cite{key-26} used no-regret learning to solve this problem, assuming that the users' channel selections are common information. The learning converges to a correlated equilibrium \cite{key-18}, wherein the commonly observed history serves as a signal to coordinate all users' channel selections. When users' channel selections are not observable, authors in \cite{key-27,key-28,key-29} designed multi-agent multi-armed bandit learning algorithms to minimize the expected performance loss of distributed spectrum access.

\begin{figure}[t]
\centering
\includegraphics[scale=0.7]{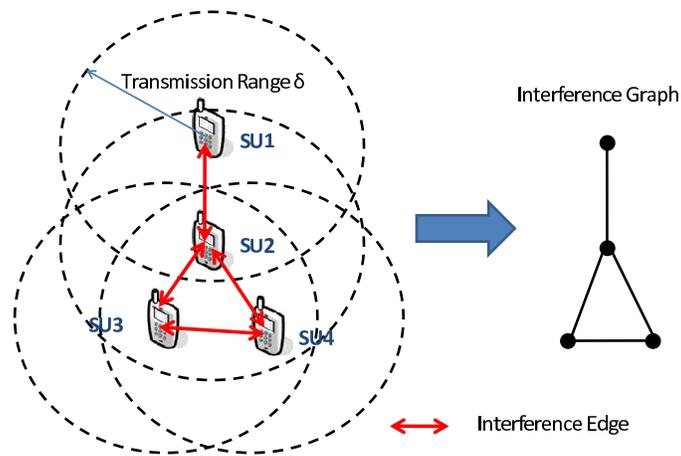}
\caption{\label{fig:Distributed-spectrum-access}Illustration of distributed
spectrum access with spatial reuse}
\end{figure}

A common assumption of the above results is that secondary users are
close-by and interfere with each other when they transmit on the same
channel simultaneously. However, a critical feature of spectrum sharing
in wireless communication is \emph{spatial reuse}. If wireless users are located sufficiently far apart, then they can transmit in the same frequency band simultaneously without causing any performance degradation (see Figure \ref{fig:Distributed-spectrum-access} for an illustration). Such spatial effect on distributed spectrum sharing
is less understood than many other aspects in existing literature \cite{key-60}, which motivates
this study.

Recently, Tekin \emph{et al.} in \cite{key-67} and Southwell \emph{et al.} in \cite{key-68} proposed a novel spatial
congestion game framework to take spatial relationship into account.
The key idea is to extend the classical congestion game upon a general undirected graph, by
assuming that a player's payoff depends on the number of its neighbors that choose the same resource (i.e., users are homogeneous in terms of channel contention). The homogeneous assumption follows from the set up of the classical congestion game (which only works on a fully connected graph).  The application of such a homogeneous model, however, is quite restricted, since users typically have heterogenous channel contention probabilities in wireless systems. For example, users of heterogeneous  wireless channel conditions may have heterogeneous  packet transmission error rates, which in turn result in heterogeneous channel contention window sizes at the equilibrium according to the distributed coordination function (DCF) of IEEE 802.11 networks \cite{han2008backoff}. This implies that users would have heterogeneous channel contention probabilities if they have heterogeneous equilibrium contention window sizes. As another example, users running heterogeneous applications would have heterogeneous channel access priorities according to the enhanced distributed channel access (EDCA) mechanism of IEEE 802.11e networks \cite{kong2004performance}.  In this paper, we extend the spatial congestion
game framework to formulate the random access based distributed spectrum
sharing problem with spatial reuse, by taking users' heterogeneous channel contention probabilities into account. Such extension is highly non-trivial, and significantly expands possible applications of the model. Moreover, we propose distributed algorithms to achieve Nash equilibria of the generalized spatial games.

We consider two game models in this paper. In the first model, secondary users have fixed spectrum access locations, and each user selects a channel to maximize its own utility in a distributed manner. We model the problem as
a spatial channel selection game. In the second more general model, users
are mobile, and they are capable to select channels and spectrum
access locations simultaneously in order to better exploit the gain of spatial reuse. We formulate the problem as a joint
spatial channel selection and mobility game. The main results and
contributions of this paper are as follows:
\begin{itemize}
\item \emph{General game formulation}: We formulate the spatial channel selection problem and the joint channel and location selection problem as noncooperative games on general interference graphs, with heterogeneous channel available data rates depending on user and location.
\item \emph{Existence of Nash equilibrium and finite improvement property}:
For both the spatial channel selection game and the joint spatial
channel selection and mobility game, we show that they are potential games,
and hence they always have at least one Nash equilibrium and possess the finite improvement
property.
\item \emph{Distributed algorithms for achieving Nash equilibrium}: For the spatial
channel selection game, we propose a distributed learning algorithm,
which globally converges
to a Nash equilibrium by only utilizing users' local observations. For the spatial channel selection and mobility
game, we propose a distributed strategic mobility algorithm, which also converges to a Nash
equilibrium, when jointly used with the distributed learning algorithm.
\item \emph{Superior performance: }Numerical results show that the Nash equilibria achieved by the proposed algorithms have only less than $8\%$ performance loss, compared
with the centralized optimal solutions.
\end{itemize}

The rest of the paper is organized as follows. We introduce the system
model and the spatial channel selection game in Sections \ref{sec:System-Model} and \ref{sec:Spatial-Channel-Selection}, respectively. We present the distributed
learning mechanism for spatial channel selection in Section \ref{sec:Distributed-Learning-Mechanism}. Then we introduce the joint
spatial channel selection and mobility game in Section \ref{sec:Joint-Channel-Selection}, and study the uniqueness and efficiency of Nash equilibrium in Section \ref{POA}.
We illustrate the performance of the proposed mechanisms through numerical
results in Section \ref{sec:Numerical-Results}, and finally conclude
in Section \ref{sec:Conclusion}. 
\section{\label{sec:System-Model}System Model}

We consider a dynamic spectrum sharing network with a set  $\mathcal{M}=\{1,2,...,M\}$ of independent and stochastically
heterogeneous primary channels. A set $\mathcal{N}=\{1,2,...,N\}$ of secondary users  try to access these channels in a distributed manner when the channels are not occupied by primary (licensed) transmissions.

To take the spatial relationship into account, we assume that the secondary users are located in a spatial
domain $\mathcal{\triangle}$, i.e., a finite set of possible spectrum access locations. We denote $d_{n}\in\mathcal{\triangle}$ as the \textbf{location} of user $n$, and $\boldsymbol{d}=(d_{1},..,d_{N})\in\Pi\triangleq\mathcal{\triangle}^{N}$
as \textbf{location profile} of all users. Each secondary user has a \textbf{transmission range} $\delta$. Then
given the location profile $\boldsymbol{d}$ of all users, we can obtain the \textbf{interference
graph} $G_{\boldsymbol{d}}=\{\mathcal{N},\mathcal{E}_{\boldsymbol{d}}\}$ to describe the interference relationship among users (see Figure \ref{fig:Distributed-spectrum-access} for an example). Here vertex set $\mathcal{N}$
is the secondary user set, and edge set $\mathcal{E}_{\boldsymbol{d}}=\{(i,j):||d_{i},d_{j}||\leq\delta,\forall i,j\neq i\in\mathcal{N}\}$
is the set of interference edges (with $||d_{i},d_{j}||$ being the
distance between locations $d_{i}$ and $d_{j}$).
If there is an interference edge between two secondary users, then they cannot successfully transmit their data on the same idle
channel simultaneously due to collision. In the sequel, we also denote the set of interfering users
with user $n$ (i.e., user $n$'s ``neighbors") under the location profile $\boldsymbol{d}$ as $\mathcal{N}_{n}(\boldsymbol{d})=\{i:(n,i)\in\mathcal{E}_{\boldsymbol{d}},i\in\mathcal{N}\}$.

\begin{figure}[tt]
\begin{center}
\includegraphics[scale=0.7]{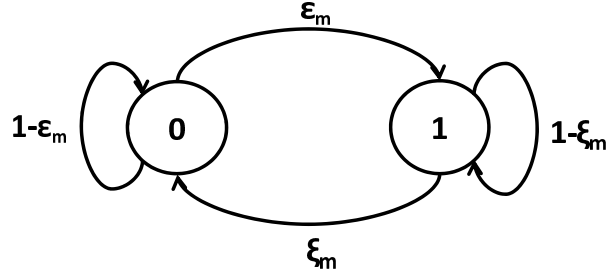}
\caption{\label{fig:Markovian-Channel-Model}Two states Markovian channel model}
\end{center}
\end{figure}

We consider a time-slotted system model as follows:
\begin{itemize}
\item \emph{Channel state}: for each primary channel $m$, the channel state
at time slot $t$ is \[
S_{m}(t)=\begin{cases}
0 & \mbox{if channel $m$ is occupied by primary transmissions,}\\
1 & \mbox{if channel $m$ is idle.}\end{cases}\]

\item \emph{Channel state changing}: the state of a channel changes according to a two-state Markovian process \cite{Zhao2007,kim2008efficient} (see Figure
\ref{fig:Markovian-Channel-Model}). We denote the channel state probability
vector of channel $m$ at time $t$ as $
\boldsymbol{q}_{m}(t)\triangleq(Pr\{S_{m}(t)=0\},Pr\{S_{m}(t)=1\}),$
 which forms a Markov chain as $
\boldsymbol{q}_{m}(t)=\boldsymbol{q}_{m}(t-1)\Gamma_{m},\forall t\geq 1,$
 with the transition matrix \[
\Gamma_{m}=\left[\begin{array}{cc}
1-\varepsilon_{m} & \varepsilon_{m}\\
\xi_{m} & 1-\xi_{m}\end{array}\right].\]

Furthermore, the long run statistical channel availability $\theta_{m}\in(0,1)$ of a channel $m$ can be obtained from the stationary distribution
of the Markov chain, i.e., \begin{align}
\theta_{m}=\frac{\varepsilon_{m}}{\varepsilon_{m}+\xi_{m}}.\label{eq:sd-2}\end{align}
\item \emph{User-and-location specific} \emph{channel throughput}: for each secondary
user $n$ at location $d$, its realized data rate $b_{m,d}^{n}(t)$
on an idle channel $m$ in each time slot $t$ evolves according
to an i.i.d. random process with a mean $B_{m,d}^{n}$, due to users'
heterogeneous transmission technologies and the local environmental
effects such as fading \cite{rappaport1996wireless}. For example, we can compute the data rate
$b_{m,d}^{n}(t)$ according to the Shannon capacity as\begin{equation}
b_{m,d}^{n}(t)=B_{m}\log_{2}\left(1+\frac{\zeta_{n}g_{m,d}^{n}(t)}{\omega_{m,d}^{n}}\right),\label{eq:dd}\end{equation}
where $B_{m}$ is the bandwidth of channel $m$, $\zeta_{n}$ is
the fixed transmission power adopted by user $n$ according to the requirements such as the primary user protection, $\omega_{m,d}^{n}$ denotes
the background noise power, and $g_{m,d}^{n}(t)$ is the channel gain. In a Rayleigh fading
channel environment, the channel gain $g_{m,d}^{n}(t)$ is a realization
of a random variable that follows the exponential distribution \cite{rappaport1996wireless}.
\item \emph{Time slot structure}: each secondary user $n$ executes the following
stages synchronously during each time slot:
\begin{itemize}
\item \emph{Channel sensing}: sense one of the channels based on the channel
selection decision generated at the end of previous time slot\footnote{This paper focuses on studying the spatial aspect  on distributed spectrum sharing, by assuming that users are capable of  perfect spectrum sensing. If a user has imperfect spectrum sensing, it would lead to a lower spectrum utilization for the user. For example, false-alarm mistakenly reports an idle channel as busy and hence results in a waste of spectrum opportunities. Missed detection mistakenly reports a busy channel as idle and results in a transmission collision with primary users. In this case, we can add a value say $\lambda_{n}$ into the throughput function in (\ref{eq:payoff1}), which describes the performance of user's spectrum sensing. If $\lambda_{n}=1$, the user has the perfect spectrum sharing. If $\lambda_{n}<1$, the user has the imperfect spectrum sensing. However, since the variable $\lambda_{n}$ does not depend on other secondary users' activities, the analysis in this paper is still valid.}.
\item \emph{Channel contention}: we use persistence-probability-based random access mechanism\footnote{This model can also provide useful insights for the case that the contention-window-based random access mechanism is implemented, since the persistence probability $p_{n}$ is related to the contention window size $w_{n}$ according to $p_{n}=\frac{2}{w_{n}+1}$ \cite{bianchi2000performance}. }, i.e., user $n$ contends for an idle channel with probability $p_{n}\in \varrho\triangleq(p_{\min},p_{\max}),$
where $0<p_{\min}<p_{\max}<1$ denote the minimum and maximum contention
probabilities. If multiple users contend for the same channel, a collision occurs and no user can transmit. Since each user (i.e., a wireless device) typically has limited battery, to achieve a longer expected lifetime, we limit user's channel contention in a time slot as
\begin{align}\zeta_{n}p_{n}\leq \nu_{n},\label{eq:cc}\end{align} where $\nu_{n}$ denotes the energy constraint of user $n$.
\item \emph{Data transmission}: transmit data packets if the user is the only one contending for an idle channel (i.e., no collision is detected).
\item \emph{Channel selection}: choose a channel to access next time slot
according to the distributed learning mechanism (introduced in Section \ref{sec:Distributed-Learning-Mechanism}).
\end{itemize}
\end{itemize}

Let $a_{n}\in\mathcal{M}$ be the channel selected by user $n$, $\boldsymbol{a}=(a_{1},...,a_{N})\in\Lambda\triangleq\mathcal{M}^{N}$
be the channel selection profile of all users, and $\boldsymbol{p}=(p_{1},...,p_{N})$
be the channel contention probability profile of all users. We can
then obtain the \emph{long run expected throughput} of each user $n$ choosing channel $a_{n}$ in location $d_{n}$ as\begin{equation}
Q_{n}(\boldsymbol{d},\boldsymbol{a},\boldsymbol{p})=\theta_{a_{n}}B_{a_{n},d_{n}}^{n}p_{n}\prod_{i\in\mathcal{N}_{n}^{a_{n}}(\boldsymbol{d},\boldsymbol{a})}(1-p_{i}),\label{eq:payoff1}\end{equation}
where $\mathcal{N}_{n}^{a_{n}}(\boldsymbol{d},\boldsymbol{a})\triangleq\{i:a_{i}=a_{n}\mbox{ and }i\in\mathcal{N}_{n}(\boldsymbol{d})\}$
is the set of interfering users that choose the same channel as user
$n$. To take the fairness issue into account, we consider the proportional-fair
utility \cite{harks2005utility} function in this study, i.e.,\begin{equation}
U_{n}(\boldsymbol{d},\boldsymbol{a},\boldsymbol{p})=\log Q_{n}(\boldsymbol{d},\boldsymbol{a},\boldsymbol{p}).\label{eq:payoff2}\end{equation}
Other type of utility functions such as general alpha-fairness will be considered
in a future work.

Equation (\ref{eq:payoff2}) shows that user $n$'s utility $U_{n}(\boldsymbol{d},\boldsymbol{a},\boldsymbol{p})$ is an increasing function of its contention probability $p_{n}$. This implies that, when a user is aggressive and does not care about the collisions, it can adopt the maximum possible channel contention probability $p_{n}$ satisfying the energy constraint (\ref{eq:cc}), i.e., $p_{n}=\min\{p_{\max},\frac{\nu_{n}}{\zeta_{n}}\}$. When users take the cost of collisions into account,  we can adopt the game theoretic framework  for the contention control in \cite{chen2007contention}. Furthermore, a dynamic contention control scheme is proposed in \cite{chen2007contention} that converges to a stable channel contention probability profile such that no users can further improve unilaterally. In this paper, we hence assume that the channel contention probability $p_{n}$ of each user is fixed and focus on the issues of distributed location and channel selections. For the sake of brevity, we also denote the utility of each user $n$ as $U_{n}(\boldsymbol{d},\boldsymbol{a})$, where the decision variables are location selections $\boldsymbol{d}$ and channel selections $\boldsymbol{a}$ only. Since our analysis is from the
secondary users' perspective, we will use the terms ``secondary user" and ``user" interchangeably. 

\section{\label{sec:Spatial-Channel-Selection}Spatial Channel Selection}

We first consider the case that all users' locations $\boldsymbol{d}$ are \emph{fixed}, and
each user tries to maximize its own utility by choosing a proper channel
in a distributed manner. Given other users' channel selections $a_{-n}$, the problem faced
by a user $n$ is \begin{equation}
\max_{a_{n}\in\mathcal{M}}U_{n}(\boldsymbol{d},a_{n},a_{-n}),\forall n\in\mathcal{N}.\label{eq:h1}\end{equation}
The distributed nature of the spatial channel selection problem naturally leads to a formulation based on game theory, such that each user
can self organize into a mutually acceptable channel selection (\textbf{Nash
equilibrium}) $\boldsymbol{a}^{*}=(a_{1}^{*},a_{2}^{*},...,a_{N}^{*})$ with \begin{equation}
a_{n}^{*}=\arg\max_{a_{n}\in\mathcal{M}}U_{n}(\boldsymbol{d},a_{n},a_{-n}^{*}),\forall n\in\mathcal{N}.\label{eq:h2}\end{equation}

We next formulate the spatial channel selection problem as a game,
and further show the existence of Nash equilibrium.

\subsection{Spatial Congestion Game}
We first review the spatial congestion game introduced in \cite{key-67}. Spatial congestion games are a class of
strategic games represented  by $\Gamma=(\mathcal{N},\mathcal{M},\{\mathcal{N}_{n}(\boldsymbol{d})\}_{n\in\mathcal{N}},\{U_{n}\}_{n\in\mathcal{N}})$.
Specifically, $\mathcal{N}$ is the set of players, $\mathcal{M}$
is the set of resources, and $\mathcal{N}_{n}(\boldsymbol{d})$ is the set of players that
can cause congestion to player $n$ when they use the same resource.
The payoff of player $n$ for using resource $a_{n}\in\mathcal{M}$
is $U_{n}(\boldsymbol{a})=f_{a_{n}}^{n}(C_{a_{n}}^{n}(\boldsymbol{a}))$, where $C_{a_{n}}^{n}(\boldsymbol{a})=\sum_{i\in\mathcal{N}_{n}(\boldsymbol{d})}I_{\{a_{i}=a_{n}\}}$
denotes the number of players in the set $\mathcal{N}_{n}(\boldsymbol{d})$ that choose
the same resource $a_{n}$ as user $n$, and $f_{a_{n}}^{n}(\cdot)$
denotes some user-specific payoff function. Typically, $C_{a_{n}}^{n}(\boldsymbol{a})$ is also called the \emph{congestion level}.

Note that the classical congestion games can be viewed as a special
case of the spatial congestion games by setting the interference graph
$G_{\boldsymbol{d}}$ as a complete graph, i.e., $\mathcal{N}_{n}(\boldsymbol{d})=\mathcal{N}\backslash\{n\}$.
For the classical congestion game, it is shown in \cite{key-41} that it
is an (exact) potential game, which is defined as
\begin{defn}[\textbf{Potential Game} \!\!\cite{key-41}]
A game is called a weighted potential game if it admits
a potential function $\Phi(\boldsymbol{a})$ such that for every $n\in\mathcal{N}$
and $a_{-n}\in\mathcal{M}^{N-1}$,\[
\Phi(a_{n}^{'},a_{-n})-\Phi(a_{n},a_{-n})=w_{n}\left(U_{n}(a_{n}^{'},a_{-n})-U_{n}(a_{n},a_{-n})\right),\]
where $w_{n}>0$ is some positive constant. Specifically, if $w_{n}=1,\forall n\in\mathcal{N}$,
then the game is also called an exact potential game. \end{defn}

\begin{defn}[\textbf{Better Response Update} \!\!\cite{key-41}]
The event where a player $n$ changes to an action $a_{n}^{'}$ from the action $a_{n}$ is a better response update if and only if $U_{n}(a_{n}^{'},a_{-n})>U_{n}(a_{n},a_{-n})$.
\end{defn}

\begin{defn}[\textbf{Finite Improvement Property} \!\!\cite{key-41}]
A game has the finite improvement property if any asynchronous better response
update process (i.e., no more than one player updates the strategy at any given
time) terminates at a pure Nash equilibrium within a finite number of updates.
\end{defn}

An appealing property of the potential game is that it admits the finite improvement property, which guarantees the existence of a Nash equilibrium. When a general payoff function $f_{a_{n}}^{n}(\cdot)$ is considered, however, the spatial congestion game does not necessarily
possess such a nice property \cite{key-67}. We next extend the spatial congestion game framework for the random access mechanism in Section \ref{sec:System-Model}, and show that the spatial channel
selection problem in (\ref{eq:h1}) with the payoff function given in (\ref{eq:payoff2}) is a potential game.

\subsection{Generalized Spatial Congestion Game Formulation}

As mentioned, the spatial congestion game proposed in \cite{key-67} assumes that a player's utility depends on the number of
players in its neighbors that choose the same resource. For our case, however, a user's utility in (\ref{eq:payoff2}) depends on who (instead
of how many users) in its neighbors contend for the same channel, since users have heterogenous channel contention probabilities. We hence generalize the spatial congestion game framework for the random access mechanism in Section \ref{sec:System-Model} by extending the definition of congestion level $C_{a_{n}}^{n}(\boldsymbol{a})$. According to (\ref{eq:payoff1}) and (\ref{eq:payoff2}),
we have\begin{align*}
U_{n}(\boldsymbol{d},\boldsymbol{a})  = \log\left(\theta_{a_{n}}B_{a_{n},d_{n}}^{n}p_{n}\right)+\sum_{i\in\mathcal{N}_{n}^{a_{n}}(\boldsymbol{d},\boldsymbol{a})}\log(1-p_{i}).\end{align*}
We then extend the definition of $C_{a_{n}}^{n}(\boldsymbol{a})$ in the standard
spatial congestion game by setting $C_{a_{n}}^{n}(\boldsymbol{a})=\sum_{i\in\mathcal{N}_{n}^{a_{n}}(\boldsymbol{d},\boldsymbol{a})}\log(1-p_{i})$.
Here $C_{a_{n}}^{n}(\boldsymbol{a})$ is regarded as the generalized congestion
level perceived by user $n$ on channel $a_{n}$. When all users have the same channel contention probability $p_{i}=p$, we have $C_{a_{n}}^{n}(\boldsymbol{a})=\log(1-p)\sum_{i\in\mathcal{N}_{n}^{a_{n}}(\boldsymbol{d},\boldsymbol{a})}I_{\{a_{i}=a_{n}\}}$, which degrades to the standard case. Then the user specific payoff function is $f_{a_{n}}^{n}(C_{a_{n}}^{n}(\boldsymbol{a}))=\log\left(\theta_{a_{n}}B_{a_{n},d_{n}}^{n}p_{n}\right)+C_{a_{n}}^{n}(\boldsymbol{a})$. In the following,
we refer to this game formulation as the spatial channel selection
game. We show that \begin{lem}
\label{lem:The-spatial-spectrum}The spatial channel selection game
on a general interference graph $G_{\boldsymbol{d}}$ is a weighted potential game,
with the potential function as\begin{align}
\Phi(\boldsymbol{d},\boldsymbol{a})=\sum_{i=1}^{N}-\log(1-p_{i})\left(\frac{1}{2}\sum_{j\in\mathcal{N}_{i}^{a_{i}}(\boldsymbol{d},\boldsymbol{a})}\log(1-p_{j})+\log\left(\theta_{a_{i}}B_{a_{i},d_{i}}^{i}p_{i}\right)\right),\label{eq:potential1}\end{align}
and the weight $w_{i}=-\log(1-p_{i})$.\end{lem}

The proof is given in Appendix \ref{proof1}. It follows from Lemma \ref{lem:The-spatial-spectrum} that
\begin{thm}
The spatial channel selection game on a general interference graph
has a Nash equilibrium and the finite improvement property.
\end{thm}
By the finite improvement property, any asynchronous better response
update leads to a Nash equilibrium. However, the better response
update requires each user to know the strategies of other users, and
then takes a better strategy to improve its payoff. This requires extensive
information exchange among the users. The signaling overhead and energy
consumption can be quite significant and even infeasible in some network
scenarios. We next propose a distributed learning mechanism, which
utilizes user's local observations only and converges to a Nash equilibrium.

\section{\label{sec:Distributed-Learning-Mechanism}Distributed Learning Mechanism
For Spatial Channel Selection}
In this part, we introduce the distributed learning mechanism for spatial channel selection, and then show that it converges to a Nash equilibrium.

\subsection{Distributed Learning Mechanism}\label{sec:dlm}

\begin{figure}[t]
\centering
\includegraphics[scale=1.0]{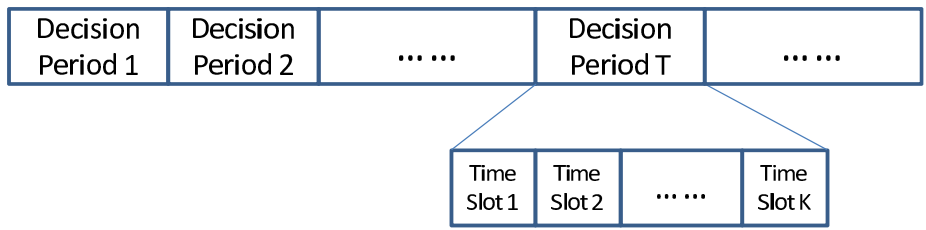}
\caption{\label{fig:Decision}Time structure of distributed learning}
\end{figure}

Without information exchange, each user can only estimate the environment
through local measurement. To achieve accurate estimation, a user
needs to gather a large number of observation samples. This motivates
us to divide the learning time into a sequence of decision periods
indexed by $T(=1,2,...)$, where each decision period consists of
$K$ time slots (see Figure \ref{fig:Decision}). During a single decision period, a user accesses
the same channel in all $K$ time slots. Thus the total number of
users accessing each channel does not change within a decision period,
which allows users to better learn the environment.

The key idea of distributed learning here is to adapt a user's spectrum access decision based on its accumulated experiences. At the beginning of each period $T$, a user $n$ chooses a channel
$a_{n}(T)\in\mathcal{M}$ to access according to its mixed strategy
$\boldsymbol{\sigma}_{n}(T)=(\sigma_{m}^{n}(T),\forall m\in\mathcal{M})$,
where $\sigma_{m}^{n}(T)$ is the probability of choosing channel
$m$. The mixed strategy is generated according to $\boldsymbol{Z}_{n}(T)=(Z_{m}^{n}(T),\forall m\in\mathcal{M}),$
which represents its perceptions of choosing different channels based
on local estimations. We map from the perceptions $\boldsymbol{Z}_{n}(T)$
to the mixed strategy $\boldsymbol{\sigma}_{n}(T)$ in the \emph{proportional}
way, i.e.,\begin{equation}
\sigma_{m}^{n}(T)=\frac{Z_{m}^{n}(T)}{\sum_{i=1}^{M}Z_{i}^{n}(T)},\forall m\in\mathcal{M}.\label{eq:DL-0}\end{equation}

At the end of a decision period $T$, a user $n$ computes its estimated
expected payoff $U_{n}(T)$ based on the sample average estimation
over $K$ time slots in the period, i.e., $U_{n}(T)=\frac{\sum_{t=1}^{K}U_{n}(T,t)}{K}$ where $U_{n}(T,t)$ is the payoff received by user $n$ in time slot $t$. Then user $n$ adjusts its perceptions
as \begin{equation}
Z_{m}^{n}(T+1)=\frac{Z_{m}^{n}(T)}{\sum_{i=1}^{M}Z_{i}^{n}(T)}+\mu_{T}U_{n}(T)I_{\{a_{n}(T)=m\}},\forall m\in\mathcal{M},\label{eq:DL-1}\end{equation}
where $\mu_{T}$ is the smoothing factor and $I_{\{a_{n}(T)=m\}}$
is an indicator whether user $n$ chooses channel $m$ at period $T$.
The user first normalizes the perception values (the first term on RHS of (\ref{eq:DL-1})) and then reinforces
the perception of the channel just accessed (the second term on RHS of (\ref{eq:DL-1})). The purpose of normalization here is to bound the perception values. We summarize the distributed
learning mechanism for spatial channel selection in Algorithm \ref{alg:Distributed-Learning-For}.

{We then analyze the computational complexity of the distributed learning algorithm. For each iteration of each user, Line $5$ involves $M$ division operations in (\ref{eq:DL-0}). This
step has the complexity of $\mathcal{O}(M)$. Similarly, Line $11$ has the complexity of $\mathcal{O}(M)$. Lines $6$ to $9$ involves $K$ channel contention in $K$ time slots and hence have the complexity of $\mathcal{O}(K)$. Line $10$ involves $K$ summation operations, which also has the complexity of $\mathcal{O}(K)$. Suppose that it takes $C$ iterations for the algorithm to converge. Then total computational complexity of the distributed learning algorithm of $N$ users is $\mathcal{O}(CNK+CNM)$.}

\begin{algorithm}[tt]
\begin{algorithmic}[1]
\State \textbf{initialization:}
\State \hspace{0.4cm} \textbf{set} the initial perception value $\boldsymbol{Z}_{n}(1)=(\frac{1}{M},...,\frac{1}{M})$.
\State \textbf{end initialization\newline}

\Loop{ for each decision period $T$ and each user $n$ in parallel:}
        \State \textbf{select} a channel $a_{n}(T)\in\mathcal{M}$ according to the mixed strategy $\boldsymbol{\sigma}_{n}(T)$ by (\ref{eq:DL-0}).
        \For{each time slot $t$ in the period $T$}
            \State \textbf{sense} and \textbf{contend} to access the channel $a_{n}(T)$.
            \State \textbf{record} the realized utility $U_{n}(T,t)$
        \EndFor
        \State \textbf{calculate} the average utility $U_{n}(T)=\frac{\sum_{t=1}^{K}U_{n}(T,t)}{K}$.
        \State \textbf{update}  the perception values $\boldsymbol{Z}_{n}(T)$ according to (\ref{eq:DL-1}).
\EndLoop

\end{algorithmic}
\caption{\label{alg:Distributed-Learning-For}Distributed Learning For Spatial Channel Selection}
\end{algorithm}

\subsection{Dynamics of Distributed Learning}

We then study the dynamics of distributed learning mechanism, which
provide useful insights for the convergence of the learning mechanism.

First of all, it is easy to show that learning procedures in (\ref{eq:DL-0}) and (\ref{eq:DL-1})
correspond to the following discrete time dynamics.
\begin{lem}
For the distributed learning mechanism for spatial channel selection, the
discrete time dynamics are given as \begin{equation}
\sigma_{m}^{n}(T+1)=\sigma_{m}^{n}(T)+\frac{\mu_{T}U_{n}(T)(I_{\{a_{n}(T)=m\}}-\sigma_{m}^{n}(T))}{1+\mu_{T}U_{n}(T)},\forall m\in\mathcal{M},n\in\mathcal{N}.\label{eq:ddynamc0}\end{equation}
\end{lem}

Since the updated perception value $Z_{m}^{n}(T)$ depends on the estimated payoff $U_{n}(T)$, $Z_{m}^{n}(T)$ is thus a random variable. The equations in (\ref{eq:ddynamc0}) are hence
stochastic difference equations, which are difficult to analyze directly.
Based on the stochastic approximation theory \cite{key-8},
we then focus on the analysis of its mean dynamics, which has the
same convergence equilibrium as the discrete dynamics (\ref{eq:ddynamc0}).

To proceed, we define the mapping from the mixed strategies $\boldsymbol{\sigma}(T)$
to the expected payoff of user $n$ choosing channel $m$ as $V_{m}^{n}(\boldsymbol{\sigma}(T))\triangleq E[U_{n}(T)|\boldsymbol{\sigma}(T),a_{n}(T)=m]$.
Here the expectation $E[\cdot]$ is taken with respect to the mixed
strategy profile $\boldsymbol{\sigma}(T)$ of all users. We show that
\begin{lem} \label{lll}
For the distributed learning mechanism for spatial channel selection, when
smoothing factor $\mu_{T}$ satisfies $\sum_{T}\mu_{T}=\infty$ and $\sum_{T}\mu_{T}^{2}<\infty$,
then as $T$ goes to infinity, the sequence $\{\boldsymbol{\sigma}(T),\forall T\geq0\}$
converges to the limiting point of the differential equations \begin{equation}
\frac{d\sigma_{m}^{n}(T)}{dT}=\sigma_{m}^{n}(T)\left(V_{m}^{n}(\boldsymbol{\sigma}(T))-\sum_{i=1}^{M}\sigma_{i}^{n}(T)V_{i}^{n}(\boldsymbol{\sigma}(T))\right),\forall m\in\mathcal{M},n\in\mathcal{N}.\label{eq:mdynamics0}\end{equation}
\end{lem}

The proof is given in Appendix \ref{proof1-2}. The mean dynamics in (\ref{eq:mdynamics0}) imply that for a user if a channel offers a
better payoff  than his current average payoff, then the user will choose that channel with a higher probability in future learning.

\subsection{Convergence of Distributed Learning}

We now study the convergence of the mean dynamics in (\ref{eq:mdynamics0}).
To proceed, we first define the following functions\begin{equation}
L(\boldsymbol{\sigma}(T)) \triangleq E[\Phi(\boldsymbol{d},\boldsymbol{a})|\boldsymbol{\sigma}(T)],\label{eq:lya1}\end{equation}
and \begin{equation}
L_{i}^{n}(\boldsymbol{\sigma}(T)) \triangleq E[\Phi(\boldsymbol{d},\boldsymbol{a})|\boldsymbol{\sigma}(T),a_{n}(T)=i].\label{eq:lay2}\end{equation}
Here $L(\boldsymbol{\sigma}(T))$ is the expected value of the potential
function $\Phi$ given the mixed strategy profile $\boldsymbol{\sigma}(T)$,
and $L_{i}^{n}(\boldsymbol{\sigma}(T))$ is the expected value of $\Phi$ given that user $n$ chooses channel
$n$ and other users adhere to the mixed strategy profile $\boldsymbol{\sigma}(T)$.
We show that
\begin{lem}\label{lem4.3} $L_{i}^{n}(\boldsymbol{\sigma}(T))-L_{j}^{n}(\boldsymbol{\sigma}(T))=-\log(1-p_{n})\left(V_{i}^{n}(\boldsymbol{\sigma}(T))-V_{j}^{n}(\boldsymbol{\sigma}(T))\right),\forall i,j\in\mathcal{M},n\in\mathcal{N}.$ \end{lem}

The proof is given in Appendix \ref{proof1-4}. This lemma implies that the potential function  of the spatial channel selection game in (\ref{eq:potential1}) also holds in the expectation sense.  Based on Lemma \ref{lem4.3}, we show that
\begin{thm}
\label{thm:For-the-distributed}When smoothing factor $\mu_{T}$ satisfies
$\sum_{T}\mu_{T}=\infty$ and $\sum_{T}\mu_{T}^{2}<\infty$, the distributed
learning mechanism for spatial channel selection asymptotically converges
to a Nash equilibrium.\end{thm}

The proof is given in Appendix \ref{proof1-3}. The key idea is to show that the time derivative of $L(\boldsymbol{\sigma}(T))$ is non-decreasing, i.e., $\frac{dL(\boldsymbol{\sigma}(T))}{dT}\geq0$. Since $L(\boldsymbol{\sigma}(T))$ is bounded above, the learning dynamics must converge to an invariant set such that $\frac{dL(\boldsymbol{\sigma}(T))}{dT}=0$, which corresponds to the set of Nash equilibria.

\section{\label{sec:Joint-Channel-Selection}Joint Spatial Channel Selection
and Mobility}

Future mobile devices are envisioned to incorporate the intelligent functionality
and will be capable of flexible spectrum access \cite{key-90}. Most
existing efforts (e.g., \cite{key-3,key-21,key-4,key-22,key-16,key-25,key-26,key-27,key-18,key-28,key-29}), however, focus on spectrum sharing
networks with stationary secondary users. How to better utilize the gain of spatial reuse in mobile cognitive radio networks is less understood. Due to the heterogeneous geo-locations of primary users, the spectrum availabilities can be very different over the spatial dimension. A secondary user can achieve higher throughput if it moves to a location with higher spectrum opportunities and fewer contending users. This motivates us to consider the throughput-driven mobility case that each user has the flexibility to change both its spectrum access location and channel.

\rev{We note that the idea of strategic mobility is not necessarily applicable to all communication scenarios.  For example, in vehicular ad-hoc networks, user's mobility is typically generated by user's driving plan, thus the idea of strategic mobility for better network throughput may not apply. However, there are some networking scenarios where strategic mobility can be very useful.  For example, in areas of poor connectivity, cellular phone users often try to find a location with better connectivity by moving around and observing the signal strength bars. As another example, in many large academic conferences, a user often experiences poor Wi-Fi connections in a conference room with a lot of attendees. The connection gets much better when the user moves into the conference lobby just tens of meters away with much fewer users. To summarize, a user has the incentive to move if he has to complete an urgent communication task and the movement is within a reasonable distance.

Strategic mobility has also been discussed in several related literature. Satyanarayanan in \cite{satyanarayanan2001pervasive} has proposed the strategic mobility for better network service as an important function of pervasive computing. An envisioned scenario is that a software agent can intelligently gather information from both the network and user and provide appropriate suggestions about location changing to the user so that the user can achieve a better communication performance.  Inspired by this, Balachandran \emph{et al.} in \cite{balachandran2002hot} proposed a network-directed roaming approach to relieve congestion in public area wireless access point networks. When an access point (i.e., a location) is over-loaded, the feedback about where to move to get less-loaded access points will be provided to users.   However, this approach computes the feedback in a centralized manner from the perspective of the network, and does not take the selfish nature of users into account. For example, it is possible that most users would choose to move to the same closest access point, which would also cause serve congestion at the new access point. We note that the strategic mobility game model in our paper works in a distributed fashion from the perspective of each individual user. For example, each user can first inform its software agent about the set of preferable candidate locations. Then all the software agents can apply the proposed algorithm to identify a mutually acceptable location selection profile for all the users (i.e., Nash equilibrium of the game).

Our proposed algorithm is also relevant to the vision of having networks of mobile agents (e.g., robots) autonomously performing sensing and communication tasks \cite{goldenberg2004towards}.  One critical issue of these networks is how to utilize the strategic mobility to improve communication performance \cite{goldenberg2004towards}.  For example, wireless mobile camera sensors with the purpose of reporting a static target to a data sink can improve their reporting data rates (i.e., achieving a higher video streaming quality) by moving strategically among the feasible locations subject to the geographical constraints of the reporting tasks.  The strategic mobility game solution in this paper, which requires no information exchange among the sensors for negotiating the location selections, can be very useful for designing a self-organizing system for such a scenario.}

\subsection{\label{sub:Strategic-Mobility-Game}Strategic Mobility Game with
Fixed Channel Selection}
We first study the case that the channel selection profile of all
users is \emph{fixed}, and users try to choose proper spectrum access
locations to maximize their own payoffs in a distributed manner. Without loss of generality,
we assume that the locations on the spatial domain $\mathcal{\triangle}$ are connected\footnote{For the case that the spatial domain is not connected,
it can be partitioned into multiple connected sub-domains.},
i.e., it is possible to get to any other locations from any location. We further introduce the user specific location selection space $\mathcal{\triangle}_{n}\subseteq\mathcal{\triangle}$ to characterize user heterogeneity in mobility preference. For example, $\mathcal{\triangle}_{n}\subseteq\mathcal{\triangle}$ can be the set of preferable candidate locations input by user $n$ to its software agent in the context of pervasive computing. If a user $n$ is willing to move all possible locations, then we have $\mathcal{\triangle}_{n}=\mathcal{\triangle}$. If the user does not want to move, we have $\mathcal{\triangle}_{n}=\{d_{n}\}$ where $d_{n}$ is user $n$'s fixed location. As another example, $\mathcal{\triangle}_{n}$ is the set of feasible locations to move subject to the geographical constraint of sensor $n$'s sensing tasks in the context of mobile sensor networks.  We then introduce the \emph{strategic mobility game} $\Omega=(\mathcal{N},\boldsymbol{d},\{U_{n}\}_{n\in\mathcal{N}})$, where $\mathcal{N}$ is the set of users,
$\boldsymbol{d}=(d_{1},...d_{n})\in\Theta\triangleq\mathcal{\triangle}_{1}\times...\times\mathcal{\triangle}_{N}$ is the location profile of all users, and $U_{n}(\boldsymbol{d},\boldsymbol{a})$
is the payoff of user $n$ given the fixed channel selection profile
$\boldsymbol{a}$ of all users. A location profile $\boldsymbol{d}^{*}=(d_{n}^{*},d_{-n}^{*})$ is a Nash equilibrium under a fixed $\boldsymbol{a}$ if and only if it satisfies that\begin{equation}
d_{n}^{*}=\arg\max_{d_{n}\in\mathcal{\triangle}_{n}}U_{n}(d_{n},d_{-n}^{*},\boldsymbol{a}),\forall n\in\mathcal{N}.\label{eq:h22}\end{equation} We show that
\begin{lem}\label{lem:hh}
The strategic mobility game $\Omega$ is a weighted potential game,
with the same potential function as $\Phi(\boldsymbol{d},\boldsymbol{a})$ in (\ref{eq:potential1}),
and the weight $w_{n}=-\log(1-p_{n})$. \end{lem}

The proof is given in Appendix \ref{proof2}. According to the property of the potential game, it follows that
\begin{thm}
The strategic mobility game $\Omega$ has a Nash equilibrium and the finite improvement property.
\end{thm}

Similarly to the spatial channel selection game, we can apply the distributed
learning mechanism to achieve the Nash equilibrium. However, due to
the cost of long distance traveling, it is often the case that each
user only desires to move to a new location that is close enough to
its current location in each single location update decision. For example, subject to the topological constraint, mobile sensors may can only move to a neighboring location in each single location update. Thus, we next propose a distributed strategic mobility
algorithm that takes this local learning constraint into consideration.

\subsection{Distributed Strategic Mobility Algorithm}

We assume that each user has a traveling distance constraint $\vartheta_{n}$,
i.e., user $n$ at location $d_{n}$ can only move to a new location
in the restricted set of locations $\mathcal{\triangle}_{d_{n}}^{n}\triangleq\{d\in\mathcal{\triangle}_{n}\backslash\{d_{n}\}:||d,d_{n}||\leq\vartheta_{n}\}$. When a user has a large enough traveling distance constraint $\vartheta_{n}$ (e.g., $\vartheta_{n}\geq\max_{d\in\mathcal{\triangle}_{n}\backslash\{d_{n}\}}\{||d,d_{n}||\}$), we will have $\mathcal{\triangle}_{d_{n}}^{n}=\mathcal{\triangle}_{n}\backslash\{d_{n}\}$ and the user would like to explore all other locations in each location update decision. When the traveling distance constraint is very small (e.g., $\vartheta_{n}=0$), then we have $\mathcal{\triangle}_{d_{n}}^{n}=\varnothing$ and the user $n$ does not want to change its location and will not involve the location selection procedure. Furthermore, we assume that each user $n$ only has the information of
its utility $U_{n}(\boldsymbol{d},\boldsymbol{a})$ through local measurement\footnote{Users can adopt the similar sample average estimation approach as in distributed learning mechanism in Section \ref{sec:dlm}.}.

\begin{figure}[t]
\centering
\includegraphics[scale=0.7]{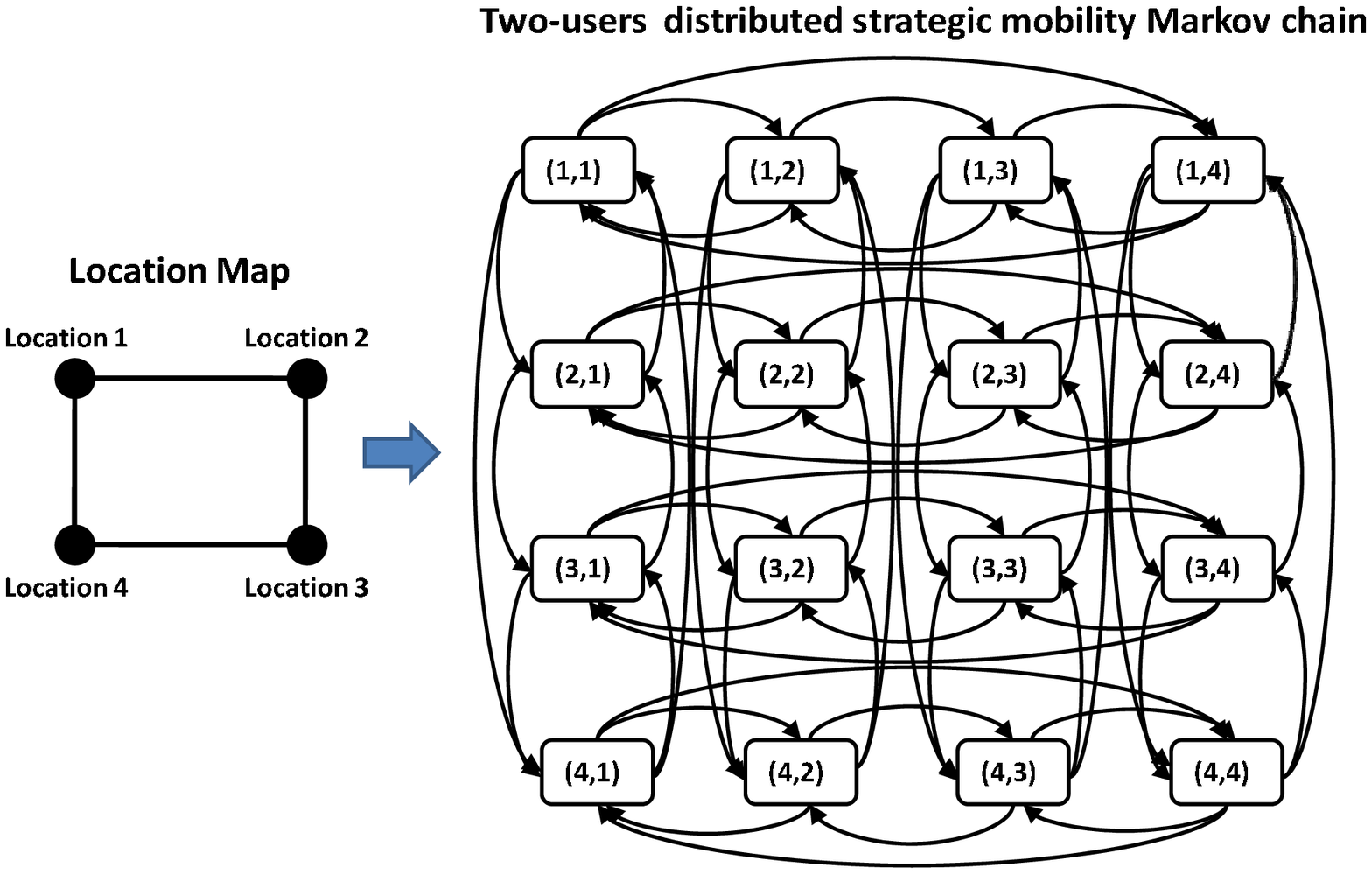}
\caption{\label{fig:Makrov}System state transition diagram of the distributed strategic mobility Markov chain by two users. In the location map on the left hand-side, one location is reachable directly from another location if these two locations are connected by an edge. In the transition diagram of the Markov chain on the right hand-side, $(d_{1},d_{2})$ denotes the system state with $d_{1}$ and $d_{2}$ being locations of user $1$ and $2$, respectively. The transition between two system states is feasible if they are connected by a link.}
\end{figure}

Motivated by the CSMA mechanism in \cite{key-31} and distributed
P2P streaming algorithm in \cite{key-33}, we design an efficient distributed
strategic mobility algorithm by carefully coordinating users' asynchronous
location updates to form a Markov chain (with the system state as the location profile $\boldsymbol{d}$ of all users). The details
of the algorithm are given in Algorithm \ref{alg:Distributed-Strategic-Mobility}. Here users update their locations asynchronously according to a timer value that follows the exponential
distribution\footnote{For ease of exposition, we have considered a Markov chain with the count-down process following an exponential distribution. It is shown in \cite{sevast1957ergodic,chandy1977product,key-31} that the convergent stationary distribution is the same as long as the state transition process (i.e., the location update process in our case) follows a general probability distribution with the same mean as in the exponential distribution case.  This implies that the proposed mobility algorithm can be implemented in a more practical way. For example, a user can update its location with a waiting time based on the power law distribution, which is a common statistical property of many human activities \cite{vazquez2006modeling}. Since we allow user specific location update density $\tau_{n}$ in the algorithm, this further implies that the waiting time for location update can be generated by user's demand and activities (e.g., dialing a phone call and writing an email at a location), rather than by the artificial count-down process.} with a rate of $\tau_{n}|\mathcal{\triangle}_{d_{n}}^{n}|$, where the density $\tau_{n}$ describes how often a user $n$ updates its location. Users with higher QoS requirement may update its location more often (i.e., with a larger timer density), in order to achieve a higher data rate. Since the exponential
distribution has support over $(0,\infty)$ and its probability density
function is continuous, the probability that more than one
users generate the same timer value and update their locations
simultaneously equals zero. Furthermore, if a user $n$ does not want to move, we have $|\mathcal{\triangle}_{d_{n}}^{n}|=0$ and hence the user $n$ will not update its location according to the algorithm. If a user has a set of candidate locations  $\mathcal{\triangle}_{d_{n}}^{n}$ to move, it will have chances to update its location selection and hence improve its utility, which also improves the system potential $\Phi(\boldsymbol{d},\boldsymbol{a})$ by the property of potential game. In the algorithm, we will use a temperature parameter $\gamma$ to control the randomness of users' location selections.  As $\gamma$ increases, a user will choose a location of higher utility with a larger probability. As an example, the system state transition diagram of the distributed strategic mobility Markov chain by two users is shown in Figure \ref{fig:Makrov}.

\begin{algorithm}[tt]
\begin{algorithmic}[1]
\State \textbf{initialization:}
\State \hspace{0.4cm} \textbf{set} the temperature $\gamma$ and the location update density $\tau_{n}$.
\State \textbf{end initialization\newline}

\Loop{ for each user $n$ in parallel:}
        \State \textbf{generate}  a timer value following  the exponential distribution
with the mean equal to $\frac{1}{\tau_{n}|\mathcal{\triangle}_{d_{n}}^{n}|}$,
where $d_{n}$ is the current location of the user and $|\mathcal{\triangle}_{d_{n}}^{n}|$
is the number of feasible locations to move to next.
        \State \textbf{count down} until the timer expires.
        \If{ the timer expires}
            \State \textbf{record} the payoff $U_{n}(\boldsymbol{d},\boldsymbol{a})$.
            \State \textbf{choose}  a new location $d_{n}^{'}$ randomly from the set $\mathcal{\triangle}_{d_{n}}^{n}$.
            \State \textbf{move} to the new location $d_{n}^{'}$ and \textbf{record} the payoff $U_{n}(\boldsymbol{d}^{'},\boldsymbol{a})$.
            \State \textbf{stay in}  the new location $d_{n}^{'}$
with probability $\frac{e^{-\log(1-p_{n})\gamma U_{n}(\boldsymbol{d}^{'},\boldsymbol{a})}}{e^{-\log(1-p_{n})\gamma U_{n}(\boldsymbol{d},\boldsymbol{a})}+e^{-\log(1-p_{n})\gamma U_{n}(\boldsymbol{d}^{'},\boldsymbol{a})}}$
OR \textbf{move back} to the original location $d_{n}$ with probability $\frac{e^{-\log(1-p_{n})\gamma U_{n}(\boldsymbol{d},\boldsymbol{a})}}{e^{-\log(1-p_{n})\gamma U_{n}(\boldsymbol{d},\boldsymbol{a})}+e^{-\log(1-p_{n})\gamma U_{n}(\boldsymbol{d}^{'},\boldsymbol{a})}}$.
        \EndIf
\EndLoop
\end{algorithmic}
\caption{\label{alg:Distributed-Strategic-Mobility}Distributed Strategic Mobility Algorithm}
\end{algorithm}

Since user $n$ will randomly choose a new location $d_{n}^{'}\in\mathcal{\triangle}_{d_{n}}^{n}$ and
stays there with probability $\\ \frac{e^{-\log(1-p_{n})\gamma U_{n}(\boldsymbol{d}^{'},\boldsymbol{a})}}{e^{-\log(1-p_{n})\gamma U_{n}(\boldsymbol{d},\boldsymbol{a})}+e^{-\log(1-p_{n})\gamma U_{n}(\boldsymbol{d}^{'},\boldsymbol{a})}}$,
then the probability from state $\boldsymbol{d}=(d_{n},d_{-n})$ to $\boldsymbol{d}^{'}=(d'_{n},d_{-n})$ is given as $\frac{1}{|\mathcal{\triangle}_{d_{n}}^{n}|}\frac{e^{-\log(1-p_{n})\gamma U_{n}(\boldsymbol{d}^{'},\boldsymbol{a})}}{e^{-\log(1-p_{n})\gamma U_{n}(\boldsymbol{d},\boldsymbol{a})}+e^{-\log(1-p_{n})\gamma U_{n}(\boldsymbol{d}^{'},\boldsymbol{a})}}.$ Since each user $n$ revises its location according to the countdown timer mechanism with a rate of $\tau_{n}|\mathcal{\triangle}_{d_{n}}^{n}|$, hence if $d'_{n}\in\mathcal{\triangle}_{d_{n}}^{n}$, the transition rate from state $\boldsymbol{d}$ to state $\boldsymbol{d}'$ is given as
\begin{align}
q_{\boldsymbol{d},\boldsymbol{d}'}=\tau_{n}\frac{e^{-\log(1-p_{n})\gamma U_{n}(\boldsymbol{d}^{'},\boldsymbol{a})}}{e^{-\log(1-p_{n})\gamma U_{n}(\boldsymbol{d},\boldsymbol{a})}+e^{-\log(1-p_{n})\gamma U_{n}(\boldsymbol{d}^{'},\boldsymbol{a})}}.\label{eq:SD2}
\end{align}
Otherwise, we have $q_{\boldsymbol{d},\boldsymbol{d}'}=0$. We show in Lemma \ref{thm:The-distributed-strategic} that the distributed strategic mobility Markov chain is time reversible. Time reversibility means that when tracing the Markov chain backwards, the stochastic behavior of the reverse Markov chain remains the same. A nice property of a time reversible Markov chain is that it always admits a unique stationary distribution, which guarantees the convergence of the distributed
strategic mobility algorithm.
\begin{lem}
\label{thm:The-distributed-strategic}The distributed strategic mobility
algorithm induces a time-reversible Markov chain with the unique stationary
distribution\begin{equation}
Pr(\boldsymbol{d},\boldsymbol{a})=\frac{e^{\gamma\Phi(\boldsymbol{d},\boldsymbol{a})}}{\sum_{\tilde{\boldsymbol{d}}\in\Theta}e^{\gamma\Phi(\tilde{\boldsymbol{d}},\boldsymbol{a})}},\forall \boldsymbol{d}\in\Theta,\label{eq:SD}\end{equation}
where $Pr(\boldsymbol{d},\boldsymbol{a})$ is the probability that the location profile $\boldsymbol{d}$
is chosen by all users under the fixed channel selection strategy profile
$\boldsymbol{a}$.\end{lem}

The proof is given in Appendix \ref{proof3}. The key of the proof is to verify that the distribution in (\ref{eq:SD}) satisfies the detailed balance equations of the distributed strategic mobility Markov chain, i.e., $Pr(\boldsymbol{d},\boldsymbol{a})q_{\boldsymbol{d},\boldsymbol{d}'}=Pr(\boldsymbol{d}^{'},\boldsymbol{a})q_{\boldsymbol{d}^{'},\boldsymbol{d}}$. Let $\Phi^{*}(\boldsymbol{a})=\max_{\boldsymbol{d}\in\Theta}\Phi(\boldsymbol{d},\boldsymbol{a})$ be the maximum of the potential function of the game,  and $\bar{\Phi}(\boldsymbol{a})$
be the expected performance by the distributed strategic mobility
algorithm. We have
\begin{thm}\label{thmmm}
For the distributed strategic mobility algorithm, as the temperature $\gamma\rightarrow\infty$, the expected performance
$\bar{\Phi}(\boldsymbol{a})$ approaches to $\Phi^{*}(\boldsymbol{a})$, and the distributed
strategic mobility algorithm converges to a Nash equilibrium.\end{thm}
\begin{proof}
Let $P_{\boldsymbol{d}}$ be the probability that the location profile $\boldsymbol{d}$ is chosen.
It is well known that the distribution $Pr(\boldsymbol{d},\boldsymbol{a})$ in (\ref{eq:SD}) is the optimal
solution for the following maximization problem \cite{key-30}:\begin{eqnarray}
\max & \sum_{\boldsymbol{d}\in\Theta}P_{\boldsymbol{d}}\Phi(\boldsymbol{d},\boldsymbol{a})-\frac{1}{\gamma}\sum_{\boldsymbol{d}\in\Theta}P_{\boldsymbol{d}}\log P_{\boldsymbol{d}}\label{eq:hah}\\
\mbox{subject to} & \sum_{\boldsymbol{d}\in\Theta}P_{\boldsymbol{d}}=1.\nonumber \end{eqnarray}
Thus, as $\gamma\rightarrow\infty$, the problem (\ref{eq:hah}) becomes the following problem
\begin{eqnarray}
\max & \sum_{\boldsymbol{d}\in\Theta}P_{\boldsymbol{d}}\Phi(\boldsymbol{d},\boldsymbol{a})\label{eq:hah1}\\
\mbox{subject to} & \sum_{\boldsymbol{d}\in\Theta}P_{\boldsymbol{d}}=1.\nonumber \end{eqnarray}

Let $P_{\boldsymbol{d}}^{*}$ be the optimal solution to problem (\ref{eq:hah1}). We thus know that, as $\gamma\rightarrow\infty$, the stationary distribution $Pr(\boldsymbol{d},\boldsymbol{a})$ approaches to $P_{\boldsymbol{d}}^{*}$. This implies that, as $\gamma\rightarrow\infty$,
$\bar{\Phi}(\boldsymbol{a})=\sum_{\boldsymbol{d}\in\Theta}Pr(\boldsymbol{d},\boldsymbol{a})\Phi(\boldsymbol{d},\boldsymbol{a})$ approaches to $\Phi^{*}(\boldsymbol{a})=\sum_{\boldsymbol{d}\in\Theta}P_{\boldsymbol{d}}^{*}\Phi(\boldsymbol{d},\boldsymbol{a})$.
\end{proof}

Note that in practice we can only implement a finite value of the temperature $\gamma$. The value of the temperature $\gamma$ is bounded such that the potential $e^{\gamma\Phi(\boldsymbol{d},\boldsymbol{a})}$  does not exceed the range of the largest predefined real number on a personal computer. Numerical results show that the algorithm with a large enough feasible $\gamma$ can converge to a near-optimal solution such that $\bar{\Phi}(\boldsymbol{a})$ is close to $\Phi^{*}(\boldsymbol{a})$. We then consider the computational complexity of the algorithm. For each iteration of each user, Lines $4$ to $15$ only involve random value generation and subduction operation for count-down, and hence have a complexity of $\mathcal{O}(1)$.  Suppose that it takes $C$ iterations for the algorithm to converge. Then total computational complexity of $N$ users is $\mathcal{O}(CN)$.

\subsection{Joint Channel Selection and Strategic Mobility}

We now consider the case that each user has the flexibility to choose
its location and channel simultaneously. Similarly to  Section \ref{sub:Strategic-Mobility-Game},
we formulate the problem as a joint spatial channel selection and mobility
game $\Upsilon=(\mathcal{N},(\boldsymbol{d},\boldsymbol{a}), \{U_{n}\}_{n\in\mathcal{N}})$. A location and channel profile $(\boldsymbol{d}^{*},\boldsymbol{a}^{*})$ is a Nash equilibrium if and only if it satisfies that\begin{equation}
(d_{n}^{*},a_{n}^{*})=\arg\max_{d_{n}\in\mathcal{\triangle}_{n},a_{n}\in\mathcal{M}}U_{n}(d_{n},d_{-n}^{*},a_{n},a_{-n}^{*}),\forall n\in\mathcal{N}.\label{eq:h222}\end{equation} We
show that the game $\Upsilon$ is also a weighted potential game.
\begin{lem} \label{lemma8}
The joint spatial channel selection and mobility game $\Upsilon$
is a weighted potential game, with the same potential function as
$\Phi(\boldsymbol{d},\boldsymbol{a})$ in (\ref{eq:potential1}), and the weight $w_{n}=-\log(1-p_{n})$. \end{lem}
\begin{proof}
Suppose that a user $k$ changes its current location $d_{k}$ and channel
$a_{k}$ to a location $d_{k}^{'}$ and channel $a_{k}^{'}$, and
the system state changes from $(\boldsymbol{d},\boldsymbol{a})$ to $(\boldsymbol{d}^{'},\boldsymbol{a}^{'})$ accordingly. Then the
change in the potential function $\Phi$ is given as\begin{align*}
\Phi(\boldsymbol{d}^{'},\boldsymbol{a}^{'})-\Phi(\boldsymbol{d},\boldsymbol{a}) = & \Phi(\boldsymbol{d}^{'},\boldsymbol{a}^{'})-\Phi(\boldsymbol{d}^{'},\boldsymbol{a})+\Phi(\boldsymbol{d}^{'},\boldsymbol{a})-\Phi(\boldsymbol{d},\boldsymbol{a})\\
= & -\log(1-p_{k})\left(U_{k}(\boldsymbol{d}^{'},\boldsymbol{a}^{'})-U_{k}(\boldsymbol{d}^{'},\boldsymbol{a})\right)-\log(1-p_{k})\left(U_{k}(\boldsymbol{d}^{'},\boldsymbol{a})-U_{k}(\boldsymbol{d},\boldsymbol{a})\right)\\
= & -\log(1-p_{k})\left(U_{k}(\boldsymbol{d}^{'},\boldsymbol{a}^{'})-U_{k}(\boldsymbol{d},\boldsymbol{a})\right), \end{align*}
which completes the proof.
\end{proof}
Lemma \ref{lemma8} implies the following key result.
\begin{thm}
The joint spatial channel selection and mobility game has a Nash equilibrium and the finite improvement property. \end{thm}
To reach a Nash equilibrium of the joint spatial channel selection and mobility game, we can
run the distributed learning mechanism for channel selection and
distributed strategic mobility algorithm together. According to
the numerical results, the distributed learning mechanism can converge
to a Nash equilibrium in less than one minute ($<$$300\times100$
time slots, and each time slot is assumed to be $2$ milliseconds,
which is longer than one normal time-slot in the standard GSM
system). Thus, we can implement the distributed strategic mobility
algorithm at a larger time-scale (say every few minutes), and implement
the distributed learning for channel selection at a smaller time scale (say every few milliseconds). Under such separation of time scales, it is reasonable to assume that  the distributed learning mechanism operating at the small time scale achieves convergence  between two updates at the large time scale. We
show that
\begin{thm} \label{thm:joint}
With the separation of time-scales, the joint distributed learning
mechanism and strategic mobility algorithm converges to a Nash equilibrium
of the joint spatial channel selection and mobility game as the temperature
$\gamma\rightarrow\infty$. \end{thm}

The proof is given in Appendix \ref{proof4}. The key idea of the proof is that the distributed learning
mechanism globally maximizes the potential function  $\Phi(\boldsymbol{d},\boldsymbol{a})$ in decision variable $\boldsymbol{a}$ given the fixed location profile $\boldsymbol{d}$, i.e., $\max_{\boldsymbol{a}}\Phi(\boldsymbol{d},\boldsymbol{a})$.  Then the strategic mobility algorithm at the larger timescale also maximizes the potential function  $\Phi(\boldsymbol{d},\boldsymbol{a})$ in terms of decision variable $\boldsymbol{d}$ given that the channel selections are $\boldsymbol{a}^{*}_{\boldsymbol{d}}=\arg\max_{\boldsymbol{a}}\Phi(\boldsymbol{d},\boldsymbol{a})$. That is, the algorithm will converge to the equilibrium such that the best location profile $\boldsymbol{d}^{*}$ with the maximum potential $\Phi(\boldsymbol{d}^{*},\boldsymbol{a}^{*}_{\boldsymbol{d}^{*}})$ will be selected. And a maximum point to the potential function is also
a Nash equilibrium of the potential game \cite{key-41}.

\section{Uniqueness and Efficiency of Nash Equilibrium}\label{POA}

In previous sections, we have considered the existence of Nash equilibrium and proposed distributed algorithms for achieving the equilibrium. We will further explore the uniqueness and efficiency of the Nash equilibrium, which can offer more useful insights for the game theoretic approach for distributed spectrum sharing with spatial reuse.

\subsection{Uniqueness of Nash equilibrium}

Due to the combinatorial nature of joint channel and location
selections, the Nash equilibrium of the game  is not unique in general.
For example, we consider a game with two users $\mathcal{N}=\{1,2\}$, two channels
$\mathcal{M}=\{1,2\}$, and two locations $\Delta=\{d^{1},d^{2}\}$.
Two locations are close such that $||d^{1},d^{2}||\leq\delta$, and
both users and channels are homogeneous such that $\theta_{m}=\theta,B_{m,d_{n}}^{n}=B,p_{n}=p$.
In this case, there are eight Nash equilibria $((d_{1},d_{2}),(a_{1},a_{2}))$ for the game, i.e.,
$\left(\left(d^{1},d^{1}\right),\left(1,2\right)\right)$, $\left(\left(d^{2},d^{2}\right),\left(1,2\right)\right)$,
$\left(\left(d^{1},d^{1}\right),\left(2,1\right)\right)$, $\left(\left(d^{2},d^{2}\right),\left(2,1\right)\right)$,$\left(\left(d^{1},d^{2}\right),\left(1,2\right)\right)$,
$\left(\left(d^{1},d^{2}\right),\left(2,1\right)\right)$,
$\left(\left(d^{2},d^{1}\right),\left(1,2\right)\right)$, and $\left(\left(d^{2},d^{1}\right),\left(2,1\right)\right)$.

In general, selecting from multiple Nash equilibria is quite hard, and the proposed algorithm is guaranteed to converge to one of the Nash equilibria.

\subsection{Price of Anarchy}

Since the Nash equilibrium is typically not unique, we then study the efficiency of Nash equilibria. Following the
definition of price of anarchy (PoA) in game theory \cite{key-16}, we will quantify
the efficiency ratio of the worst-case Nash equilibrium over the centralized optimal solution. We first consider the spatial channel selection game with
a fixed spectrum access location profile $\boldsymbol{d}$. Let $\Xi$
be the set of Nash equilibria of the game. Then the PoA is defined
as\[
\mbox{PoA}=\frac{\min_{\boldsymbol{a}\in\Xi}\sum_{n\in\mathcal{N}}U_{n}(\boldsymbol{d},\boldsymbol{a})}{\max_{\boldsymbol{a}\in\mathcal{M}^{N}}\sum_{n\in\mathcal{N}}U_{n}(\boldsymbol{d},\boldsymbol{a})},\]
which is always not greater than $1$. A larger PoA implies that the set of Nash equilibrium is more efficient (in the worst-case sense) using the centralized optimum as a benchmark. Let $\varpi=\max_{n\in\mathcal{N}}\{-\log(1-p_{n})\}$,
$E(\boldsymbol{d})=\min_{n\in\mathcal{N}}\max_{m\in\mathcal{M}}\left\{ \log\left(\theta_{m}B_{m,d_{n}}^{n}p_{n}\right)\right\} $,
and $K(\boldsymbol{d})=\max_{n\in\mathcal{N}}\{|\mathcal{N}_{n}(\boldsymbol{d})|\}$. We can show that
\begin{thm}
\label{thm:For-the-spatialPoA}For the spatial channel selection game
with a fixed spectrum access location profile $\boldsymbol{d}$, the
PoA is no less than $1-\frac{K(\boldsymbol{d})\varpi}{E(\boldsymbol{d})}$.\end{thm}

The proof is given in Appendix \ref{proofPoA}. Intuitively, when users are less aggressive in channel contention (i.e., $\varpi$ is smaller) and users are more homogeneous in term of channel utilization (i.e., $E(\boldsymbol{d})$ is larger), the worst-case Nash equilibrium is closer to the centralized optimum and hence the PoA is larger.  Moreover, Theorem \ref{thm:For-the-spatialPoA} implies that we can increase the efficiency of spectrum sharing  by better utilizing the gain of spatial reuse (i.e., reducing the interference edges
$K(\boldsymbol{d})$ on the interference graph). Similarly, by defining
that $\eta=\max_{\boldsymbol{d}\in\Theta}\{\frac{K(\boldsymbol{d})}{E(\boldsymbol{d})}\}$,
we see that the PoA of the joint spatial channel selection and mobility game is no less than $1-\eta\varpi$.

The PoA characterizes the worst-case performance of Nash equilibria. Numerical results in Section \ref{sec:Numerical-Results} demonstrate that the convergent Nash equilibrium of the proposed algorithm is often more efficient and has a less than $8\%$ performance loss,
compared with the centralized optimal solution.

\section{\label{sec:Numerical-Results}Numerical Results}

We now evaluate the proposed algorithms by simulations. We consider
a Rayleigh fading channel environment. The data rate of secondary user $n$
on an idle channel $m$ at location $d$ is given as $b_{m,d}^{n}=h_{d}b_{m}^{n}.$
Here $h_{d}$ is a location dependent parameter. Parameter $b_{m}^{n}$
is the data rate computed according to the Shannon capacity, i.e., $b_{m}^{n}=B_{m}\log_{2}(1+\frac{\zeta_{n}g_{m}^{n}}{\omega_{m,d}^{n}})$,
where $B_{m}$ is the bandwidth of channel $m$, $\zeta_{n}$ is the
power adopted by user $n$, $\omega_{m,d}^{n}$ is the noise power, and
$g_{m}^{n}$ is the channel gain (a realization of a random variable that
follows the exponential distribution with the mean $\bar{g}_{m}^{n}$).
In the following simulations, we set $B_{m}=10$ MHz, $\omega_{m,d}^{n}=-100$
dBm, and $\zeta_{n}=100$ mW. By choosing different location parameter $h_{d}$
and mean channel gain $\bar{g}_{m}^{n}$, we have different mean data
rates $E[b_{m,d}^{n}]=B_{m,d}^{n}=h_{d}E[b_{m}^{n}]=h_{d}B_{m}^{n}$
for different channels, locations, and users. For simplicity, we set
the channel availabilities $\theta_{m}=0.5$.

\begin{figure}[t]
\centering
\includegraphics[scale=0.6]{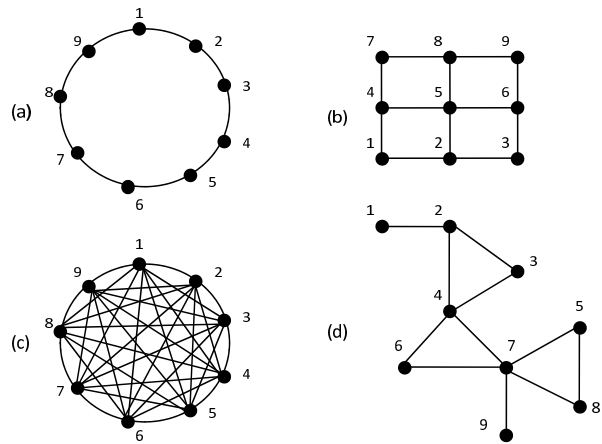}
\caption{\label{fig:Interference-Graphs}Interference graphs \vspace{0.8cm}}
\end{figure}

\subsection{\label{channelselection}Distributed Learning For Spatial Channel Selection}


We first evaluate the distributed learning algorithm for channel selection
with fixed user locations. For the distributed learning algorithm
initialization, we set the length of each decision period $K=100$,
which can achieve a good estimation of the expected payoff. For the smoothing factor $\mu_{T}$, a higher value can lead to a faster convergence. We hence set
$\mu_{T}=\frac{1}{T}$, which has the fastest convergence while satisfying the convergence condition in Theorem \ref{thm:For-the-distributed}.

Since locations are fixed, we set the location parameter $h_{d}=1$. We
consider a network of $M=5$ channels and $N=9$ users with four different
interference graphs (see Figure \ref{fig:Interference-Graphs}). Graphs (a), (b) and (c) are the commonly-used regular interference graphs, and Graph (d) is a randomly-generated non-regular interference graph.  Let $\vec{B}_{n}=(B_{1}^{n},...,B_{M}^{n})$ be the mean data rate vector of user $n$ on $M$ channels. We set $\vec{B}_{1}=\vec{B}_{2}=\vec{B}_{3}=(0.1,0.3,0.8,1.0,1.5)$
Mbps, $\vec{B}_{4}=\vec{B}_{5}=\vec{B}_{6}=(0.2,0.6,1.6,2.0,3.0)$
Mbps, and $\vec{B}_{7}=\vec{B}_{8}=\vec{B}_{9}=(0.5,1.5,4.0,5.0,7.5)$
Mbps. The fixed channel contention probabilities $p_{n}$ of the users are randomly assigned from the set $\{0.1,0.2,...,0.9\}$.

\begin{figure}[t]
\centering
\includegraphics[scale=0.6]{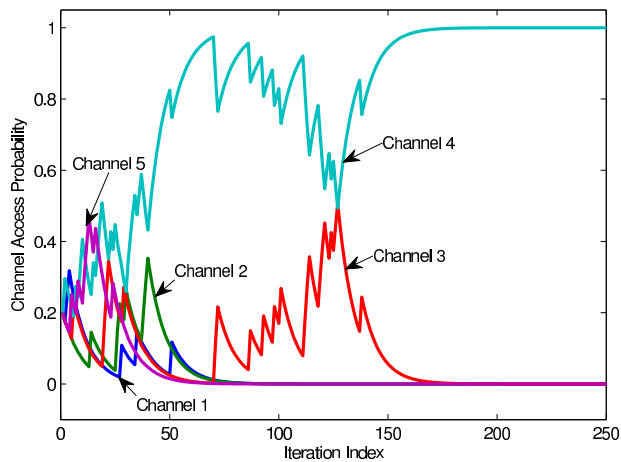}
\caption{\label{fig:Learning-dynamics-of-1}Learning dynamics of user 4's channel
selection probabilities \newline \vspace{0.8cm}}
\end{figure}

\begin{figure}[t]
\centering
\includegraphics[scale=0.6]{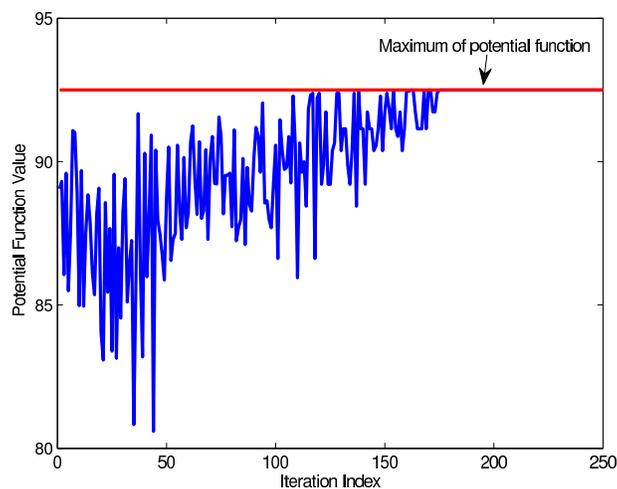}
\caption{\label{fig:Learning-dynamics-of2}Learning dynamics of potential function value \newline \vspace{0.8cm}}
\end{figure}

Let us first look at the convergence dynamics, using graph (d) in Figure \ref{fig:Interference-Graphs} as an example. Figure \ref{fig:Learning-dynamics-of-1} shows the learning dynamics of user $4$ in terms of the channel selection probabilities on $5$ channels. It demonstrates the convergence of the distributed learning algorithm. Figure \ref{fig:Learning-dynamics-of2} shows the learning dynamics of the potential function value $\Phi$.
We see that the distributed learning algorithm can lead the potential
function of the spatial channel selection game to the maximum
point, which is a Nash equilibrium according to the property of potential
game.

\begin{figure}[t]
\centering
\includegraphics[scale=1.2]{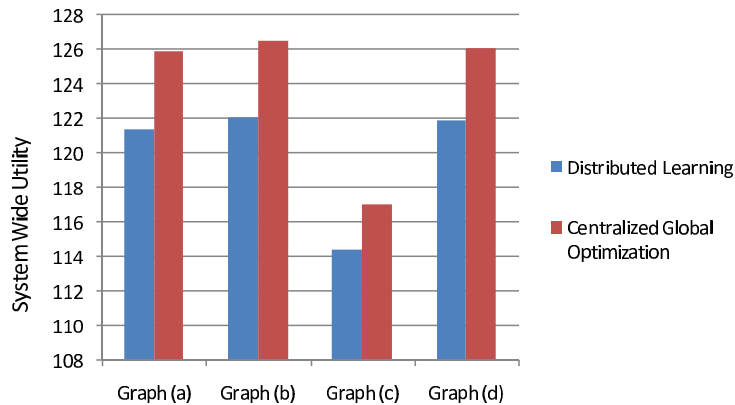}
\caption{\label{fig:Comparison-of-the}Comparison of distributed learning and
global optimization \vspace{0.8cm}}
\end{figure}

To benchmark the performance of the distributed learning algorithm,
we compare it with the solution obtained by the centralized global
optimization of $\max_{\boldsymbol{a}}\sum_{n\in\mathcal{N}}U_{n}(\boldsymbol{d},\boldsymbol{a})$ on all
the interference graphs. The results are shown in Figure \ref{fig:Comparison-of-the}.
We see that the performance loss of the distributed
learning is less than $5\%$ in all cases.

\begin{figure}[t]
\centering
\includegraphics[scale=0.5]{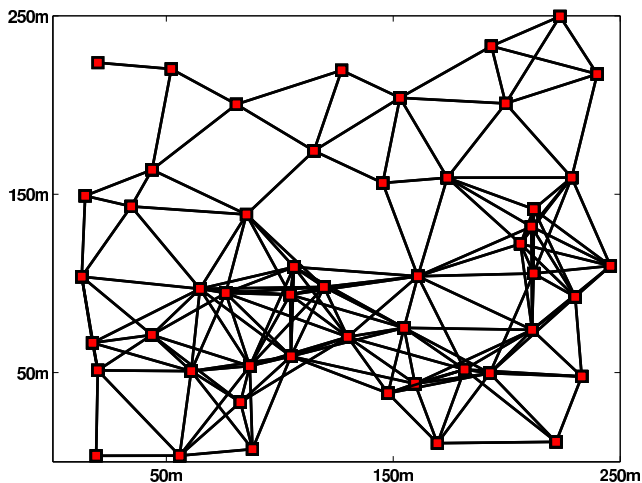}
\caption{\label{fig:50Nodes}A square area of a length of $250$m with 50 scattered
users with an transmission range $\delta=60$m. Each user is represented by a dot and two users interfere with each other if they are connected by an edge. \vspace{0.8cm}}
\end{figure}

\begin{figure}[t]
\centering
\includegraphics[scale=1.2]{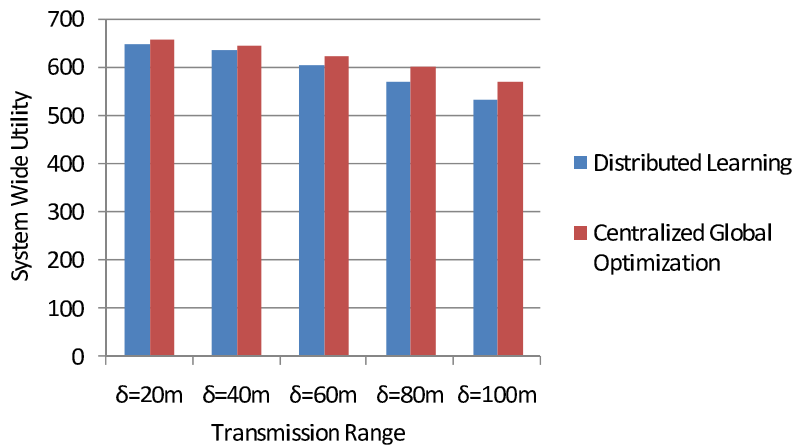}
\caption{\label{fig:50NodesComparison}Comparison of distributed learning and
global optimization with different transmission ranges $\delta$ \vspace{0.8cm}}
\end{figure}

We look at another network with $N=50$ users randomly scattered across a square area of a side-length of $250$m (see Figure \ref{fig:50Nodes}). We set users' transmission range $\delta=20, 40, 60, 80,$ and $100$m, respectively. Figure \ref{fig:50NodesComparison} shows the performance comparison between distributed learning and the centralized optimization solution. As the transmission range $\delta$ increases, the performances of both distributed learning and  centralized global
optimization solutions decrease. In all cases, the performance loss of the distributed learning algorithm is less than $8\%$, compared with the centralized global optimization solution.   This shows the  efficiency of distributed learning algorithm.

\subsection{Joint Distributed Learning and Strategic Mobility}

\begin{figure}[t]
\centering
\includegraphics[scale=0.6]{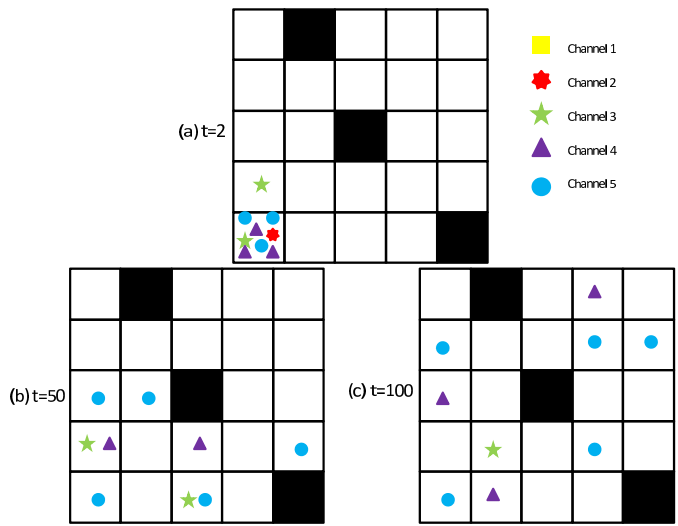}
\caption{\label{fig:Dynamics-of-joint}Dynamics of users' locations and channel
selections with the temperature $\gamma=50$ \newline}
\end{figure}

We next study the joint distributed learning and strategic mobility
algorithm. We consider a location map as shown in graph (a) of Figure \ref{fig:Dynamics-of-joint}.
Black cells are obstacles, and no users can move there. Each user
in a cell can interfere with those users within the same
cell and the ones in neighboring cells (along the line and diagonal).
Each user initially locates in the same cell in the bottom left corner, and is allowed to move to the neighboring cells
once it gets the chance to update its location. Each
cell is randomly assigned with a location parameter $h_{d}$ from the
set $\{0.5,1.0,2.0\}$, and each user has different mean data rates
$B_{m}^{n}$ as specified in Section \ref{channelselection}.

\begin{figure}[t]
\centering
\includegraphics[scale=0.5]{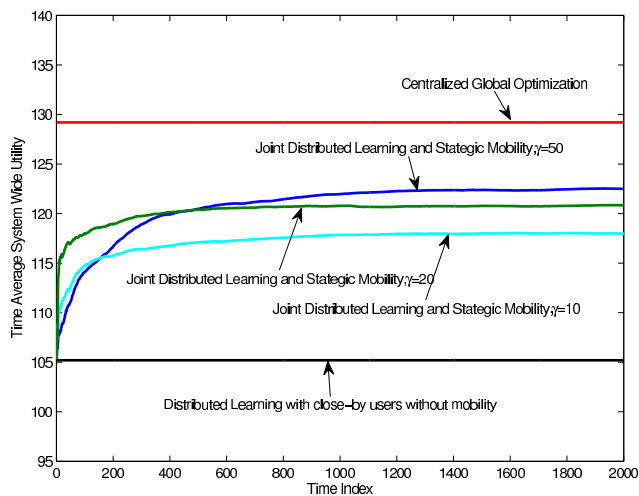}
\caption{\label{fig:Dynamics-of-time2}Dynamics of time average system utility with location update process following the exponential distribution and the temperature $\gamma=10,$ $20$, and $50$, respectively}
\end{figure}

We implement the joint algorithm with the temperature $\gamma=10,$ $20$, and $50$, respectively. The location update process follows the exponential distribution with a mean of $10$. We show in Figure \ref{fig:Dynamics-of-joint} users' locations and channel selections
at the iteration step $t=2,50$, and $100$, respectively (with the temperature $\gamma=50$). We observe that users try to spread out in terms of physical locations and meanwhile choose channels with higher data rates, in order to maximize their payoffs. From Figure \ref{fig:Dynamics-of-time2}, we see
that the performance of the algorithm improves as the temperature
$\gamma$ increases, and the convergence time also increases accordingly.
When $\gamma=50$, the performance loss of the joint algorithm is less than
$6\%$, compared with the global optimal solution, i.e., $\max_{\boldsymbol{d},\boldsymbol{a}}\sum_{n\in\mathcal{N}}U_{n}(\boldsymbol{d},\boldsymbol{a})$. This shows the efficiency of the Nash equilibrium. When users are static (without strategic mobility) and close-by, the performance
loss of the distributed learning for channel selection can be as high
as $18\%$, which justifies the motivations for the strategic mobility
design.

\begin{figure}[t]
\centering
\includegraphics[scale=0.5]{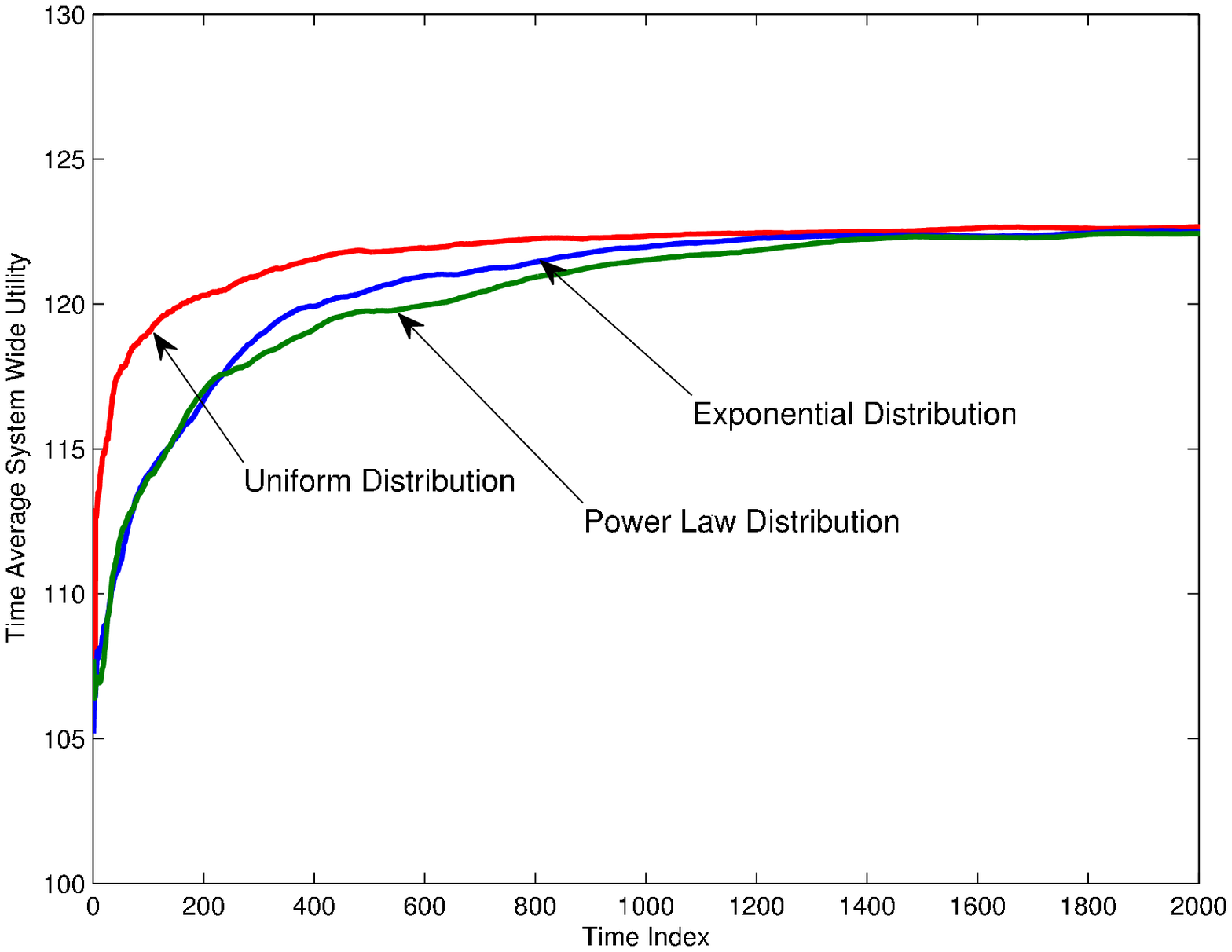}
\caption{\label{fig:Distributions}Dynamics of time average system utility with the location update process following different distributions and the temperature $\gamma=50$}
\end{figure}

\rev{We further implement simulations where the temperature $\gamma=50$ and the location update process follows the  uniform distribution and power law distribution with the same mean as in the exponential distribution case, respectively. The results in Figure \ref{fig:Distributions}  verify that the convergent system performance is the same as long as the location update process follows a general probability distribution with the same mean as in the exponential distribution case. Moreover, we observe that the convergence time increases when the distribution has a longer tail (e.g., power law distribution). This is because that a small fraction of users would have a longer waiting time for the location update when a long-tailed distribution is implemented.}

\section{\label{sec:Conclusion}Conclusion}
In this paper, we generalize the spatial
congestion game framework for  distributed spectrum
access mechanism design with spatial reuse. We consider both the spatial channel
selection game and the joint spatial channel selection and mobility game, and propose distributed algorithms using users' local information that converge to the Nash equilibria for both games. Numerical results verify that Nash equilibria are quite efficient and have less than $8\%$ performance loss, compared with the centralized optimal solutions.

For the future work, we are going to investigate the distributed spectrum sharing mechanism design with spatial reuse that can achieve the centralized optimal solution. 
\appendix\label{appendixA}
\subsection{Proof of Lemma \ref{lem:The-spatial-spectrum}}\label{proof1}
For the ease of exposition, we first define $\rho_{i}\triangleq\log(1-p_{i})$,
$\xi_{m,d}^{i}\triangleq\log(\theta_{m}B_{m,d}^{i}p_{i})$, and\[
\Phi_{i}^{m}(\boldsymbol{d},\boldsymbol{a})=-\rho_{i}\left(\frac{1}{2}\sum_{j\in\mathcal{N}_{i}^{m}(\boldsymbol{d},\boldsymbol{a})}\rho_{j}+\xi_{m,d_{i}}^{i}\right)I_{\{a_{i}=m\}}.\]
Thus, we have $\Phi(\boldsymbol{d},\boldsymbol{a})=\sum_{i=1}^{N}\sum_{m=1}^{M}\Phi_{i}^{m}(\boldsymbol{d},\boldsymbol{a})$.

Now suppose that a user $k$ unilaterally changes its strategy $a_{k}$
to $a_{k}^{'}$. Let $\boldsymbol{a}'=(a_{1},...,a_{k-1},a_{k}^{'},a_{k+1},...,a_{N})$
be the new strategy profile. Thus, the change in potential $\Phi$
from $\boldsymbol{a}$ to $\boldsymbol{a}^{'}$ is given by\begin{align}
 & \Phi(\boldsymbol{d},\boldsymbol{a}^{'})-\Phi(\boldsymbol{d},\boldsymbol{a})= \sum_{i=1}^{N}\sum_{m=1}^{M}\Phi_{i}^{m}(\boldsymbol{d},\boldsymbol{a}^{'})-\sum_{i=1}^{N}\sum_{m=1}^{M}\Phi_{i}^{m}(\boldsymbol{d},\boldsymbol{a})\nonumber \\
= & \sum_{m=1}^{M}\Phi_{k}^{m}(\boldsymbol{d},\boldsymbol{a}^{'})-\sum_{m=1}^{M}\Phi_{k}^{m}(\boldsymbol{d},\boldsymbol{a})+\sum_{i\in\mathcal{N}_{k}(\boldsymbol{d})}\sum_{m=1}^{M}\Phi_{i}^{m}(\boldsymbol{d},\boldsymbol{a}^{'})-\sum_{i\in\mathcal{N}_{k}(\boldsymbol{d})}\sum_{m=1}^{M}\Phi_{i}^{m}(\boldsymbol{d},\boldsymbol{a})\nonumber \\
= & \left(\sum_{m=1}^{M}\Phi_{k}^{m}(\boldsymbol{d},\boldsymbol{a}^{'})-\sum_{m=1}^{M}\Phi_{k}^{m}(\boldsymbol{d},\boldsymbol{a})\right)+\sum_{i\in\mathcal{N}_{k}(\boldsymbol{d})}\left(\Phi_{i}^{a_{k}^{'}}(\boldsymbol{d},\boldsymbol{a}^{'})-\Phi_{i}^{a_{k}^{'}}(\boldsymbol{d},\boldsymbol{a})\right)+\sum_{i\in\mathcal{N}_{k}(\boldsymbol{d})}\left(\Phi_{i}^{a_{k}}(\boldsymbol{d},\boldsymbol{a}^{'})-\Phi_{i}^{a_{k}}(\boldsymbol{d},\boldsymbol{a})\right).\label{eq:P1}\end{align}

Equation (\ref{eq:P1}) consists of three parts. Next we analyze each part separately. For the first part, we have\begin{align}
 & \sum_{m=1}^{M}\Phi_{k}^{m}(\boldsymbol{d},\boldsymbol{a}^{'})-\sum_{m=1}^{M}\Phi_{k}^{m}(\boldsymbol{d},\boldsymbol{a})\nonumber \\
= & \Phi_{k}^{a_{k}^{'}}(\boldsymbol{d},\boldsymbol{a}^{'})-\Phi_{k}^{a_{k}}(\boldsymbol{d},\boldsymbol{a}) = -\rho_{k}\left(\frac{1}{2}\sum_{j\in\mathcal{N}_{k}^{a_{k}^{'}}(\boldsymbol{d},\boldsymbol{a}^{'})}\rho_{j}+\xi_{a_{k}^{'},d_{k}}^{k}\right)+\rho_{k}\left(\frac{1}{2}\sum_{j\in\mathcal{N}_{k}^{a_{k}}(\boldsymbol{d},\boldsymbol{a})}\rho_{j}+\xi_{a_{k},d_{k}}^{k}\right).\label{eq:P2}\end{align}

For the second part in (\ref{eq:P1}), \begin{align*}
 & \Phi_{i}^{a_{k}^{'}}(\boldsymbol{d},\boldsymbol{a}^{'})-\Phi_{i}^{a_{k}^{'}}(\boldsymbol{d},\boldsymbol{a})\\
= & -\rho_{i}\left(\frac{1}{2}\sum_{j\in\mathcal{N}_{i}^{a_{k}^{'}}(\boldsymbol{d},\boldsymbol{a}^{'})}\rho_{j}+\xi_{a_{k}^{'},d_{i}}^{i}\right)I_{\{a_{i}=a_{k}^{'}\}}+\rho_{i}\left(\frac{1}{2}\sum_{j\in\mathcal{N}_{i}^{a_{k}^{'}}(\boldsymbol{d},\boldsymbol{a})}\rho_{j}+\xi_{a_{k}^{'},d_{i}}^{i}\right)I_{\{a_{i}=a_{k}^{'}\}}\\
= & -\frac{1}{2}\rho_{i}\left(\sum_{j\in\mathcal{N}_{i}^{a_{k}^{'}}(\boldsymbol{d},\boldsymbol{a}^{'})}\rho_{j}-\sum_{j\in\mathcal{N}_{i}^{a_{k}^{'}}(\boldsymbol{d},\boldsymbol{a})}\rho_{j}\right)I_{\{a_{i}=a_{k}^{'}\}}= -\frac{1}{2}\rho_{i}\rho_{k}I_{\{a_{i}=a_{k}^{'}\}}.\end{align*}
This means \begin{align}
 \sum_{i\in\mathcal{N}_{k}(\boldsymbol{d})}\left(\Phi_{i}^{a_{k}^{'}}(\boldsymbol{d},\boldsymbol{a}^{'})-\Phi_{i}^{a_{k}^{'}}(\boldsymbol{d},\boldsymbol{a})\right)=\sum_{i\in\mathcal{N}_{k}(\boldsymbol{d})}-\frac{1}{2}\rho_{i}\rho_{k}I_{\{a_{i}=a_{k}^{'}\}}=-\frac{1}{2}\rho_{k}\sum_{i\in\mathcal{N}_{k}^{a_{k}^{'}}(\boldsymbol{d},\boldsymbol{a}^{'})}\rho_{i}.\label{eq:P4}\end{align}
For the third term in (\ref{eq:P1}), we can similarly get\begin{align}
 & \sum_{i\in\mathcal{N}_{k}(\boldsymbol{d})}\left(\Phi_{i}^{a_{k}}(\boldsymbol{d},\boldsymbol{a}^{'})-\Phi_{i}^{a_{k}}(\boldsymbol{d},\boldsymbol{a})\right)=\frac{1}{2}\rho_{k}\sum_{i\in\mathcal{N}_{k}^{a_{k}^{'}}(\boldsymbol{d},\boldsymbol{a})}\rho_{i}.\label{eq:P5-1}\end{align}
Substituting (\ref{eq:P2}), (\ref{eq:P4}), and (\ref{eq:P5-1}) into
(\ref{eq:P1}), we obtain\begin{align}
 & \Phi(\boldsymbol{d},\boldsymbol{a}^{'})-\Phi(\boldsymbol{d},\boldsymbol{a})\nonumber \\
= & -\rho_{k}\left(\sum_{j\in\mathcal{N}_{k}^{a_{k}^{'}}(\boldsymbol{d},\boldsymbol{a}^{'})}\rho_{j}+\xi_{a_{k}^{'},d_{k}}^{k}-\sum_{j\in\mathcal{N}_{k}^{a_{k}}(\boldsymbol{d},\boldsymbol{a})}\rho_{j}-\xi_{a_{k},d_{k}}^{k}\right)= -\rho_{k}\left(U_{k}(\boldsymbol{d},\boldsymbol{a}^{'})-U_{k}(\boldsymbol{d},\boldsymbol{a})\right).\label{eq:P6}\end{align}

Since $0<p_{k}<1$ and hence $-\log(1-p_{k})>0$, we can conclude
that $\Phi(\boldsymbol{d},\boldsymbol{a})$ defining in (\ref{eq:potential1}) is a weighted
potential function with the weight $-\log(1-p_{k})$.
\qed

\subsection{Proof of Lemma \ref{lll}}\label{proof1-2}
We complete the proof by checking the assumptions of Theorem 2.1 in
\cite{key-8}(pp.127).

(a) Since $0<\theta_{m},p_{n}<1$ and $b_{m,d}^{n}$ is bounded, then
$U_{n}(T)$ must be also bounded. It follows
that $|U_{n}(T)(I_{\{a_{n}(T)=m\}}-\sigma_{m}^{n}(T))|<\infty$.
Thus, $\sup_{T}E[|U_{n}(T)(I_{\{a_{n}(T)=m\}}-\sigma_{m}^{n}(T))|^{2}]<\infty$.

(b) First, we can obtain from (\ref{eq:ddynamc0}) that \begin{eqnarray}
\frac{d\sigma_{m}^{n}(T)}{dT} & = & \lim_{\mu_{T}\rightarrow0}\frac{\sigma_{m}^{n}(T+1)-\sigma_{m}^{n}(T)}{\mu_{T}}=U_{n}(T)(I_{\{a_{n}(T)=m\}}-\sigma_{m}^{n}(T)).\label{eq:mdynamics1}\end{eqnarray}
By taking the expectation of the RHS of (\ref{eq:mdynamics1}) with
respective to $\boldsymbol{\sigma}(T)$, we have\begin{align*}
 E[U_{n}(T)(I_{\{a_{n}(T)=m\}}-\sigma_{m}^{n}(T))|\boldsymbol{\sigma}(T)] = & \sigma_{m}^{n}(T)(1-\sigma_{m}^{n}(T))V_{m}^{n}(\boldsymbol{\sigma}(T))+\sum_{i\neq m}\sigma_{i}^{n}(T)(1-\sigma_{m}^{n}(T))V_{i}^{n}(\boldsymbol{\sigma}(T))\\
  = & \sigma_{m}^{n}(T)\sum_{i=1}^{M}\sigma_{i}^{n}(T)(V_{m}^{n}(\boldsymbol{\sigma}(T))-V_{i}^{n}(\boldsymbol{\sigma}(T))).\end{align*}

(c) First, $V_{m}^{n}(\boldsymbol{\sigma}(T))=E[U_{n}(T)|\boldsymbol{\sigma}(T),a_{n}(T)=m]$
is an expectation function, and hence is differentiable. It then follows
that $\sigma_{m}^{n}(T)\sum_{i=1}^{M}\sigma_{i}^{n}(T)(V_{m}^{n}(\boldsymbol{\sigma}(T))-V_{i}^{n}(\boldsymbol{\sigma}(T)))$
is also differentiable because the sum of differentiable functions
is also differentiable. Thus $\sigma_{m}^{n}(T)\sum_{i=1}^{M}\sigma_{i}^{n}(T)(V_{m}^{n}(\boldsymbol{\sigma}(T))-V_{i}^{n}(\boldsymbol{\sigma}(T)))$
is continuous.

(d) We have $\sum_{T}\mu_{T}=\infty$ and $\sum_{T}\mu_{T}^{2}<\infty$
by assumption.

(e) Since the sample average estimation is unbiased, the noise term
is hence the martingale difference noise. Then the expected biased
error $\beta_{T}=0$. It follows that $\sum_{T}\mu_{T}|\beta_{T}|<\infty$
with probability one.\qed

\subsection{Proof of Lemma \ref{lem4.3}}\label{proof1-4}
Let $\boldsymbol{a}=(i, a_{-n})$ and $\boldsymbol{a}^{'}=(j, a_{-n})$. By the definition of $V_{n}^{n}(\boldsymbol{\sigma}(T))$, we first have that\begin{align}
 & V_{i}^{n}(\boldsymbol{\sigma}(T))-V_{j}^{n}(\boldsymbol{\sigma}(T))\nonumber \\
 = & E\left[\log\left(\theta_{a_{n}}B_{a_{n},d_{n}}^{n}p_{n}\prod_{n'\in\mathcal{N}_{n}^{a_{n}}(\boldsymbol{d},a_{n},a_{-n})}(1-p_{n'})\right)|a_{n}=i,\boldsymbol{\sigma}(T)\right]\nonumber \\
 & - E\left[\log\left(\theta_{a_{n}}B_{a_{n},d_{n}}^{n}p_{n}\prod_{n'\in\mathcal{N}_{n}^{a_{n}}(\boldsymbol{d},a_{n},a_{-n})}(1-p_{n'})\right)|a_{n}=j,\boldsymbol{\sigma}(T)\right]\nonumber \\
= & E\left[\log\left(\theta_{i}B_{i,d_{n}}^{n}p_{n}\prod_{n'\in\mathcal{N}_{n}^{i}(\boldsymbol{d},\boldsymbol{a})}(1-p_{n'})\right)-\log\left(\theta_{j}B_{j,d_{n}}^{n}p_{n}\prod_{n'\in\mathcal{N}_{n}^{j}(\boldsymbol{d},\boldsymbol{a}^{'})}(1-p_{n'})\right)|\boldsymbol{\sigma}_{-n}(T)\right]\nonumber \\
= & \sum_{a_{-n}}\left(\log\theta_{i}B_{i,d_{n}}^{n}p_{n}\prod_{n'\in\mathcal{N}_{n}^{i}(\boldsymbol{d},\boldsymbol{a})}(1-p_{n'})-\log\theta_{j}B_{j,d_{n}}^{n}p_{n}\prod_{n'\in\mathcal{N}_{n}^{j}(\boldsymbol{d},\boldsymbol{a}^{'})}(1-p_{n'})\right)Pr\{a_{-n}|\boldsymbol{\sigma}_{-n}(T)\}\label{eq:l1}\end{align}
According to (\ref{eq:P6}), we have

\begin{align}
 & \log\left(\theta_{i}B_{i,d_{n}}^{n}p_{n}\prod_{n'\in\mathcal{N}_{n}^{i}(\boldsymbol{d},\boldsymbol{a})}(1-p_{n'})\right)-\log\left(\theta_{j}B_{j,d_{n}}^{n}p_{n}\prod_{n'\in\mathcal{N}_{n}^{j}(\boldsymbol{d},\boldsymbol{a}^{'})}(1-p_{n'})\right)\nonumber \\
= & \frac{1}{-\log(1-p_{n})}\sum_{k=1}^{N}-\log(1-p_{k})\left(\frac{1}{2}\sum_{n'\in\mathcal{N}_{k}^{a_{k}}(\boldsymbol{d},\boldsymbol{a})}\log(1-p_{n'})+\log\theta_{a_{k}}B_{a_{k},d_{k}}^{k}p_{k}\right)\nonumber \\
 & -\frac{1}{-\log(1-p_{n})}\sum_{k=1}^{N}-\log(1-p_{k})\left(\frac{1}{2}\sum_{n'\in\mathcal{N}_{k}^{a_{k}}(\boldsymbol{d},\boldsymbol{a}^{'})}\log(1-p_{n'})+\log\theta_{a_{k}}B_{a_{k},d_{k}}^{k}p_{k}\right).\label{eq:l2}\end{align}
By (\ref{eq:l1}) and (\ref{eq:l2}), it follows that\begin{align*}
 & -\log(1-p_{n})\left(V_{i}^{n}(\boldsymbol{\sigma}(T))-V_{j}^{n}(\boldsymbol{\sigma}(T))\right)\\
= & \sum_{a_{-n}}Pr\{a_{-n}|\boldsymbol{\sigma}_{-n}(T)\}\left(\sum_{k=1}^{N}-\log(1-p_{k})\right.\left(\frac{1}{2}\sum_{n'\in\mathcal{N}_{k}^{a_{k}}(\boldsymbol{d},\boldsymbol{a})}\log(1-p_{n'})+\log\theta_{a_{k}}B_{a_{k},d_{k}}^{k}p_{k}\right)\\
 & -\sum_{k=1}^{N}-\log(1-p_{k})\left.\left(\frac{1}{2}\sum_{n'\in\mathcal{N}_{k}^{a_{k}}(\boldsymbol{d},\boldsymbol{a}^{'})}\log(1-p_{n'})+\log\theta_{a_{k}}B_{a_{k},d_{k}}^{k}p_{k}\right)\right)\\
= & L_{i}^{n}(\boldsymbol{\sigma}(T))-L_{j}^{n}(\boldsymbol{\sigma}(T)),\end{align*}
which completes the proof. \qed

\subsection{Proof of Theorem \ref{thm:For-the-distributed}}\label{proof1-3}
We first consider the variation of $L(\boldsymbol{\sigma}(T))$ along
the trajectories of ODE in (\ref{eq:mdynamics0}), i.e., differentiating
$L(\boldsymbol{\sigma}(T))$ with respective to time $T$, \begin{align}
    & \frac{dL(\boldsymbol{\sigma}(T))}{dT}= \sum_{j=1}^{M}\frac{dL(\boldsymbol{\sigma}(T))}{d\sigma_{j}^{n}(T)}\frac{d\sigma_{j}^{n}(T)}{dT}\nonumber \\
  = & \sum_{j=1}^{M}L_{j}^{n}(\boldsymbol{\sigma}(T))\sigma_{j}^{n}(T)\sum_{i=1}^{M}\sigma_{i}^{n}(T)\left(V_{j}^{n}(\boldsymbol{\sigma}(T))-V_{i}^{n}(\boldsymbol{\sigma}(T))\right)\nonumber \\
  = & \frac{1}{2}\sum_{j=1}^{M}\sum_{i=1}^{M}\sigma_{j}^{n}(T)\sigma_{i}^{n}(T)\left(V_{j}^{n}(\boldsymbol{\sigma}(T))-V_{i}^{n}(\boldsymbol{\sigma}(T))\right)\left(L_{j}^{n}(\boldsymbol{\sigma}(T))-L_{i}^{n}(\boldsymbol{\sigma}(T))\right).\label{eq:pr4}\end{align}
According to Lemma \ref{lem4.3}, we have $\frac{dL(\boldsymbol{\sigma}(T))}{dT}\geq0.$
Hence $L(\boldsymbol{\sigma}(T))$ is non-decreasing along the trajectories
of the ODE (\ref{eq:mdynamics0}). According to \cite{key-99}, the
learning mechanism converges to a stationary point
$\boldsymbol{\sigma}^{*}$ such that $\frac{dL(\boldsymbol{\sigma}^{*})}{dT}=0,$
i.e., ($\forall i,j\in\mathcal{M},n\in\mathcal{N}$)\begin{align}
\sigma_{j}^{n*}\sigma_{i}^{n*}(V_{j}^{n}(\boldsymbol{\sigma}^{*})-V_{i}^{n}(\boldsymbol{\sigma}^{*}))(L_{j}^{n}(\boldsymbol{\sigma}^{*})-L_{i}^{n}(\boldsymbol{\sigma}^{*}))= \sigma_{j}^{n*}\sigma_{i}^{n*}(V_{j}^{n}(\boldsymbol{\sigma}^{*})-V_{i}^{n}(\boldsymbol{\sigma}^{*}))^{2}=0.\label{eq:pr8}\end{align}
According to (\ref{eq:mdynamics0}) and (\ref{eq:pr8}), we have $\frac{d\sigma_{m}^{n*}}{dT}=0,\forall m\in\mathcal{M},n\in\mathcal{N}.$

If $\boldsymbol{\sigma}^{*}$ is a Nash equilibrium, it must satisfy
that\begin{equation}
V_{i}^{n}(\boldsymbol{\sigma}^{*})\leq\sum_{j=1}^{M}\sigma_{j}^{n*}V_{j}^{n}(\boldsymbol{\sigma}^{*}),\forall n\in\mathcal{N},i\in\mathcal{M}.\label{eq:pr7}\end{equation}
If $\boldsymbol{\sigma}^{*}$ is not a Nash equilibrium, we must have
that there is some $n$ and $i$ such that $V_{i}^{n}(\boldsymbol{\sigma}^{*})>\sum_{j=1}^{M}\sigma_{j}^{n*}V_{j}^{n}(\boldsymbol{\sigma}^{*}).$
Due to the continuity of the expectation function $V_{i}^{n}$, the
inequality will still hold in a small open neighborhood around $\boldsymbol{\sigma}^{*}$.
Then it follows from (\ref{eq:mdynamics0}) that, for all points $\hat{\boldsymbol{\sigma}}$
in this neighborhood of $\boldsymbol{\sigma}^{*}$ that satisfy $\hat{\sigma}_{i}^{n}\neq0$,
we have \begin{align*}
\frac{d\hat{\sigma}_{i}^{n}}{dT}  = \hat{\sigma}_{i}^{n}(V_{i}^{n}(\hat{\boldsymbol{\sigma}})-\sum_{j=1}^{M}\hat{\sigma}_{j}^{n}V_{j}^{n}(\hat{\boldsymbol{\sigma}})) > 0.\end{align*}
Hence in all sufficiently small neighborhoods of $\boldsymbol{\sigma}^{*}$,
there will be infinitely many points starting from which $\hat{\boldsymbol{\sigma}}$
will eventually leave the neighborhood. Thus, the learning mechanism must asymptotically converge to a stable stationary
point $\boldsymbol{\sigma}^{*}$ that satisfies (\ref{eq:pr7}), which is a Nash equilibrium. Moreover, according to the Sard's theorem \cite{schoen1994lectures}, when the ODE (\ref{eq:mdynamics0}) asymptotically converges, the converging equilibrium is not contained in the interior of the mixed strategy polytope \cite{kleinberg2009multiplicative}. That is, for each user $n$, there exists only one channel selection $a_{n}^{*}\in\mathcal{M}$ such that $\sigma_{m}^{n*}=1$ if $m=a_{n}^{*}$ and $\sigma_{m}^{n*}=0$ otherwise.  The learning mechanism hence converges to a Nash equilibrium with pure strategy profile. \qed

\subsection{Proof of Lemma \ref{lem:hh}}\label{proof2}
Suppose that a user $k$ changes its location $d_{k}$ to the location
$d_{k}^{'}$. Let $\boldsymbol{d}^{'}=(d_{1},..,d_{k-1},d_{k}^{'},d_{k+1},...,d_{N})$.
Recall that $\rho_{i}=\log(1-p_{i})$ and $\xi_{m,d}^{i}=\log(\theta_{m}B_{m,d}^{i}p_{i})$  as defined in Appendix \ref{proof1}.
Then the change in potential $\Phi$ from $\boldsymbol{d}$ to $\boldsymbol{d}^{'}$ is given
by\begin{align}
\Phi(\boldsymbol{d}^{'},\boldsymbol{a})-\Phi(\boldsymbol{d},\boldsymbol{a})= -\rho_{k}(\xi_{a_{k},d_{k}^{'}}^{k}-\xi_{a_{k},d_{k}}^{k})+\sum_{i=1}^{N}-\frac{1}{2}\rho_{i}\left(\sum_{j\in\mathcal{N}_{i}^{a_{i}}(\boldsymbol{d}^{'},\boldsymbol{a})}\rho_{j}-\sum_{j\in\mathcal{N}_{i}^{a_{i}}(\boldsymbol{d},\boldsymbol{a})}\rho_{j}\right).\label{eq:sm0}\end{align}

For the last term, we have \begin{align}
 & \sum_{i=1}^{N}-\frac{1}{2}\rho_{i}\left(\sum_{j\in\mathcal{N}_{i}^{a_{i}}(\boldsymbol{d}^{'},\boldsymbol{a})}\rho_{j}-\sum_{j\in\mathcal{N}_{i}^{a_{i}}(\boldsymbol{d},\boldsymbol{a})}\rho_{j}\right)\nonumber \\
= & -\frac{1}{2}\rho_{k}\left(\sum_{j\in\mathcal{N}_{k}^{a_{k}}(\boldsymbol{d}^{'},\boldsymbol{a})}\rho_{j}-\sum_{j\in\mathcal{N}_{k}^{a_{k}}(\boldsymbol{d},\boldsymbol{a})}\rho_{j}\right)-\frac{1}{2}\sum_{n\in\mathcal{N}}I_{\{n\in\mathcal{N}_{k}(\boldsymbol{d}^{'},\boldsymbol{a})\cup\mathcal{N}_{k}(\boldsymbol{d},\boldsymbol{a})\}}\rho_{n}\left(\sum_{j\in\mathcal{N}_{n}^{a_{n}}(\boldsymbol{d}',\boldsymbol{a})}\rho_{j}-\sum_{j\in\mathcal{N}_{n}^{a_{n}}(\boldsymbol{d},\boldsymbol{a})}\rho_{j}\right)\nonumber \\
= & -\frac{1}{2}\rho_{k}\left(\sum_{j\in\mathcal{N}_{k}^{a_{k}}(\boldsymbol{d}^{'},\boldsymbol{a})}\rho_{j}-\sum_{j\in\mathcal{N}_{k}^{a_{k}}(\boldsymbol{d},\boldsymbol{a})}\rho_{j}\right)-\frac{1}{2}\sum_{n\in\mathcal{N}}I_{\{n\in\mathcal{N}_{k}(\boldsymbol{d}^{'},\boldsymbol{a})\cup\mathcal{N}_{k}(\boldsymbol{d},\boldsymbol{a})\}}\rho_{n}\rho_{k}\left(I_{\{n\in\mathcal{N}_{k}^{a_{k}}(\boldsymbol{d}',\boldsymbol{a})\}}-I_{\{n\in\mathcal{N}_{k}^{a_{k}}(\boldsymbol{d},\boldsymbol{a})\}}\right)\nonumber \\
= & -\frac{1}{2}\rho_{k}\left(\sum_{j\in\mathcal{N}_{k}^{a_{k}}(\boldsymbol{d}^{'},\boldsymbol{a})}\rho_{j}-\sum_{j\in\mathcal{N}_{k}^{a_{k}}(\boldsymbol{d},\boldsymbol{a})}\rho_{j}\right) -\frac{1}{2}\rho_{k}\sum_{n\in\mathcal{N}}I_{\{n\in\mathcal{N}_{k}(\boldsymbol{d}^{'},\boldsymbol{a})\cup\mathcal{N}_{k}(\boldsymbol{d},\boldsymbol{a})\}}I_{\{n\in\mathcal{N}_{k}^{a_{k}}(\boldsymbol{d}',\boldsymbol{a})\}}\rho_{n}\nonumber \\
& +\frac{1}{2}\rho_{k}\sum_{n\in\mathcal{N}}I_{\{n\in\mathcal{N}_{k}(\boldsymbol{d}^{'},\boldsymbol{a})\cup\mathcal{N}_{k}(\boldsymbol{d},\boldsymbol{a})\}}I_{\{n\in\mathcal{N}_{k}^{a_{k}}(\boldsymbol{d},\boldsymbol{a})\}}\rho_{n}\nonumber \\
= & -\rho_{k}\left(\sum_{j\in\mathcal{N}_{k}^{a_{k}}(\boldsymbol{d}^{'},\boldsymbol{a})}\rho_{j}-\sum_{j\in\mathcal{N}_{k}^{a_{k}}(\boldsymbol{d},\boldsymbol{a})}\rho_{j}\right).\label{eq:sm1}\end{align}
Combing (\ref{eq:sm0}) and (\ref{eq:sm1}), we have $\Phi(\boldsymbol{d}^{'},\boldsymbol{a})-\Phi(\boldsymbol{d},\boldsymbol{a})=-\log(1-p_{k})\left(U_{k}(\boldsymbol{d}^{'},\boldsymbol{a})-U_{k}(\boldsymbol{d},\boldsymbol{a})\right).$ \qed

\subsection{Proof of Lemma \ref{thm:The-distributed-strategic}} \label{proof3}
As mentioned, the system state of the distributed strategic mobility Markov chain is defined as the location profile
$\boldsymbol{d}\in\Theta$ of all users. Since distance measure is symmetry, we
have that if $d^{'}\in\mathcal{\triangle}_{d}^{n}$ then $d\in\mathcal{\triangle}_{d^{'}}^{n}$.
Further, since all locations on the spatial domain $\Delta$ are connected,
all system states $\boldsymbol{d}$ hence can reach each other within a finite
number of transitions, and the resulting finite Markov chain is irreducible
and aperiodic. The process is thus ergodic and has a unique stationary
distribution.

We then show the Markov chain is time reversible by checking the following
detailed balance equations are satisfied:\begin{equation}
Pr(\boldsymbol{d},\boldsymbol{a})q_{\boldsymbol{d},\boldsymbol{d}'}=Pr(\boldsymbol{d}^{'},\boldsymbol{a})q_{\boldsymbol{d}^{'},\boldsymbol{d}},\forall \boldsymbol{d},\boldsymbol{d}^{'}\in\Theta,\label{eq:SD1}\end{equation}
where $q_{\boldsymbol{d},\boldsymbol{d}'}$ is the transition rate from state $\boldsymbol{d}=(d_{1},...,d_{N})$
to state $\boldsymbol{d}^{'}=(d_{1}^{'},...,d_{N}^{'})$. According to the algorithm,
we know that the set of states that is directed connected to the state
$\boldsymbol{d}$ are the one where $\boldsymbol{d}$ and $\boldsymbol{d}^{'}$ differ by exactly one user,
say user $n$, such that $d_{i}=d_{i}^{'}, \forall i\neq n$ and $d_{n}\neq d_{n}^{'}$.

Since user $n$ revises its location by the timer mechanism, according the system state transition rate in (\ref{eq:SD2}), we have that\begin{align}
   Pr(\boldsymbol{d},\boldsymbol{a})q_{\boldsymbol{d},\boldsymbol{d}'} = \tau_{n}\frac{e^{\gamma\Phi(\boldsymbol{d},\boldsymbol{a})}}{\sum_{\tilde{\boldsymbol{d}}\in\mathcal{\boldsymbol{d}}^{N}}e^{\gamma\Phi(\tilde{\boldsymbol{d}},\boldsymbol{a})}}\frac{e^{-\log(1-p_{n})\gamma U_{n}(\boldsymbol{d}^{'},\boldsymbol{a})}}{e^{-\log(1-p_{n})\gamma U_{n}(\boldsymbol{d},\boldsymbol{a})}+e^{-\log(1-p_{n})\gamma U_{n}(\boldsymbol{d}^{'},\boldsymbol{a})}}.\label{eq:SD3}\end{align}
Similarly, we obtain that\begin{align}
   Pr(\boldsymbol{d},\boldsymbol{a})q_{\boldsymbol{d},\boldsymbol{d}'} = \tau_{n}\frac{e^{\gamma\Phi(\boldsymbol{d}^{'},\boldsymbol{a})}}{\sum_{\tilde{\boldsymbol{d}}\in\mathcal{\boldsymbol{d}}^{N}}e^{\gamma\Phi(\tilde{\boldsymbol{d}},\boldsymbol{a})}}\frac{e^{-\log(1-p_{n})\gamma U_{n}(\boldsymbol{d},\boldsymbol{a})}}{e^{-\log(1-p_{n})\gamma U_{n}(\boldsymbol{d},\boldsymbol{a})}+e^{-\log(1-p_{n})\gamma U_{n}(\boldsymbol{d}^{'},\boldsymbol{a})}}.\label{eq:SD4}\end{align}

Since the strategic mobility game is a potential game, we have\begin{equation}
\Phi(\boldsymbol{d}^{'},\boldsymbol{a})-\Phi(\boldsymbol{d},\boldsymbol{a})=-\log(1-p_{n})\left(U_{n}(\boldsymbol{d}^{'},\boldsymbol{a})-U_{n}(\boldsymbol{d},\boldsymbol{a})\right).\label{eq:SD5}\end{equation}

Combing (\ref{eq:SD3}), (\ref{eq:SD4}) and (\ref{eq:SD5}), we have
detailed balance equation (\ref{eq:SD1}) hold. The Markov chain
is hence time-reversible and has the stationary distribution given
in (\ref{eq:SD}). \qed

\subsection{Proof of Theorem \ref{thm:joint}}\label{proof4}
According to Theorem \ref{thm:For-the-distributed}, we know the distributed
learning algorithm can converge to the Nash equilibrium of the spatial channel
selection game. Let $\boldsymbol{a}_{\boldsymbol{d}}^{*}$ be the Nash equilibrium by the distributed
learning algorithm when the location profile of all users are $\boldsymbol{d}$.
Since the Nash equilibrium of the potential game is also a maximum
point to the potential function, we have $\boldsymbol{a}_{\boldsymbol{d}}^{*}=\arg\max_{\boldsymbol{a}}\Phi(\boldsymbol{d},\boldsymbol{a})$.


Similarly as the analysis of the distributed strategic
mobility algorithm in Lemma \ref{thm:The-distributed-strategic},
we define the system state of the joint channel selection and strategic mobility Markov chain as the location profile
$\boldsymbol{d}\in\Theta$ of all users. Since distance measure is symmetry, we
have that if $d^{'}\in\mathcal{\triangle}_{d}^{n}$ then $d\in\mathcal{\triangle}_{d^{'}}^{n}$.
Further, since all locations on the spatial domain $\Delta$ are connected,
all system states $\boldsymbol{d}$ hence can reach each other within a finite
number of transitions, and the resulting finite Markov chain is irreducible
and aperiodic.

We then show the Markov chain is time reversible by checking the following
detailed balance equations are satisfied:\begin{equation}
Pr(\boldsymbol{d})q_{\boldsymbol{d},\boldsymbol{d}'}=Pr(\boldsymbol{d}^{'})q_{\boldsymbol{d}^{'},\boldsymbol{d}},\forall \boldsymbol{d},\boldsymbol{d}^{'}\in\Theta,\label{eq:sd11}\end{equation}
where $q_{\boldsymbol{d},\boldsymbol{d}'}$ is the transition rate from state $\boldsymbol{d}=(d_{1},...,d_{N})$
to state $\boldsymbol{d}^{'}=(d_{1}^{'},...,d_{N}^{'})$. According to the algorithm,
we know that the set of states that is directed connected to the state
$\boldsymbol{d}$ are the one where $\boldsymbol{d}$ and $\boldsymbol{d}^{'}$ differ by exactly one user,
say user $n$, such that $d_{i}=d_{i}^{'}\forall i\neq n$ and $d_{n}\neq d_{n}^{'}$.
Since the user $n$ revise its location by the timer mechanism, we
know that the rate of revision is equal to $\tau_{n}|\mathcal{\triangle}_{d_{n}}^{n}|$.
Since user $n$ will randomly choose a new location $d_{n}^{'}$ and
stays there with probability $\frac{e^{-\log(1-p_{n})\gamma U_{n}(\boldsymbol{d}^{'},\boldsymbol{a}_{\boldsymbol{d}^{'}}^{*})}}{e^{-\log(1-p_{n})\gamma U_{n}(\boldsymbol{d},\boldsymbol{a}_{\boldsymbol{d}}^{*})}+e^{-\log(1-p_{n})\gamma U_{n}(\boldsymbol{d}^{'},\boldsymbol{a}_{\boldsymbol{d}^{'}}^{*})}}$,
the probability from state $\boldsymbol{d}$ to $\boldsymbol{d}^{'}$ is then given as $\frac{1}{|\mathcal{\triangle}_{d_{n}}^{n}|}\frac{e^{-\log(1-p_{n})\gamma U_{n}(\boldsymbol{d}^{'},\boldsymbol{a}_{\boldsymbol{d}^{'}}^{*})}}{e^{-\log(1-p_{n})\gamma U_{n}(\boldsymbol{d},\boldsymbol{a}_{\boldsymbol{d}}^{*})}+e^{-\log(1-p_{n})\gamma U_{n}(\boldsymbol{d}^{'},\boldsymbol{a}_{\boldsymbol{d}^{'}}^{*})}}.$
Thus ,the transition rate from state $\boldsymbol{d}$ to $\boldsymbol{d}^{'}$is give as \begin{equation}
q_{\boldsymbol{d},\boldsymbol{d}'}=\tau_{n}\frac{e^{-\log(1-p_{n})\gamma U_{n}(\boldsymbol{d}^{'},\boldsymbol{a}_{\boldsymbol{d}^{'}}^{*})}}{e^{-\log(1-p_{n})\gamma U_{n}(\boldsymbol{d},\boldsymbol{a}_{\boldsymbol{d}}^{*})}+e^{-\log(1-p_{n})\gamma U_{n}(\boldsymbol{d}^{'},\boldsymbol{a}_{\boldsymbol{d}^{'}}^{*})}}.\label{eq:sd12}\end{equation}
It follows that\begin{align}
 Pr(\boldsymbol{d})q_{\boldsymbol{d},\boldsymbol{d}'}= \tau_{n}\frac{e^{\gamma\Phi(\boldsymbol{d},\boldsymbol{a}_{\boldsymbol{d}}^{*})}}{\sum_{\tilde{\boldsymbol{d}}\in\mathcal{\boldsymbol{d}}^{N}}e^{\gamma\Phi(\tilde{\boldsymbol{d}},\boldsymbol{a}_{\tilde{\boldsymbol{d}}}^{*})}}\frac{e^{-\log(1-p_{n})\gamma U_{n}(\boldsymbol{d}^{'},\boldsymbol{a}_{\boldsymbol{d}^{'}}^{*})}}{e^{-\log(1-p_{n})\gamma U_{n}(\boldsymbol{d},\boldsymbol{a}_{\boldsymbol{d}}^{*})}+e^{-\log(1-p_{n})\gamma U_{n}(\boldsymbol{d}^{'},\boldsymbol{a}_{\boldsymbol{d}^{'}}^{*})}}.\label{eq:sd13} \end{align}

Similarly, we obtain that\begin{align}
  Pr(\boldsymbol{d}^{'})q_{\boldsymbol{d}^{'},\boldsymbol{d}} = \tau_{n}\frac{e^{\gamma\Phi(\boldsymbol{d}^{'},\boldsymbol{a}_{\boldsymbol{d}^{'}}^{*})}}{\sum_{\tilde{\boldsymbol{d}}\in\mathcal{\boldsymbol{d}}^{N}}e^{\gamma\Phi(\tilde{\boldsymbol{d}},\boldsymbol{a}_{\tilde{\boldsymbol{d}}}^{*})}}\frac{e^{-\log(1-p_{n})\gamma U_{n}(\boldsymbol{d},\boldsymbol{a}_{\boldsymbol{d}^{'}}^{*})}}{e^{-\log(1-p_{n})\gamma U_{n}(\boldsymbol{d},\boldsymbol{a}_{\boldsymbol{d}}^{*})}+e^{-\log(1-p_{n})\gamma U_{n}(\boldsymbol{d}^{'},\boldsymbol{a}_{\boldsymbol{d}^{'}}^{*})}}.\label{eq:14} \end{align}

Since the joint channel selection and strategic mobility game is a potential game, we have\begin{align}
 \Phi(\boldsymbol{d}^{'},\boldsymbol{a}_{\boldsymbol{d}^{'}}^{*})-\Phi(\boldsymbol{d},\boldsymbol{a}_{\boldsymbol{d}}^{*})=-\log(1-p_{n})\left(U_{n}(\boldsymbol{d}^{'},\boldsymbol{a}_{\boldsymbol{d}^{'}}^{*})-U_{n}(\boldsymbol{d},\boldsymbol{a}_{\boldsymbol{d}}^{*})\right).\label{eq:15}\end{align}

Combing (\ref{eq:sd13}), (\ref{eq:14}) and (\ref{eq:15}), we have
detailed balance equation (\ref{eq:sd11}) holds. The Markov chain
is hence time-reversible and has the unique stationary distribution given as $
Pr(\boldsymbol{d})=\frac{e^{\gamma\Phi(\boldsymbol{d},\boldsymbol{a}_{\boldsymbol{d}}^{*})}}{\sum_{\tilde{\boldsymbol{d}}\in\mathcal{\boldsymbol{d}}^{N}}e^{\gamma\Phi(\tilde{\boldsymbol{d}},\boldsymbol{a}_{\tilde{\boldsymbol{d}}}^{*})}},\boldsymbol{d}\in\Theta$.

With the similar proof as in Theorem \ref{thmmm}, we can hence show that, as $\gamma\rightarrow\infty$, the algorithm approaches
the equilibrium such that $\Phi(\boldsymbol{d},\boldsymbol{a}_{\boldsymbol{d}}^{*})$ is maximized in term of decision variable $\boldsymbol{d}$, i.e., $\max_{\boldsymbol{d}}\Phi(\boldsymbol{d},\boldsymbol{a}_{\boldsymbol{d}}^{*})$. Furthermore, we can show that $\max_{\boldsymbol{d}}\Phi(\boldsymbol{d},\boldsymbol{a}_{\boldsymbol{d}}^{*})=\max_{\boldsymbol{d},\boldsymbol{a}}\Phi(\boldsymbol{d},\boldsymbol{a})$ by contradiction. Let $\boldsymbol{d}^{*}=\arg\max_{\boldsymbol{d}}\Phi(\boldsymbol{d},\boldsymbol{a}_{\boldsymbol{d}}^{*})$ and $(\bar{\boldsymbol{d}},\bar{\boldsymbol{a}})=\arg\max_{\boldsymbol{d},\boldsymbol{a}}\Phi(\boldsymbol{d},\boldsymbol{a})$. Suppose that $\Phi(\boldsymbol{d}^{*},\boldsymbol{a}_{\boldsymbol{d}^{*}}^{*})<\Phi(\bar{\boldsymbol{d}},\bar{\boldsymbol{a}})$.
Since the learning algorithm maximizes the potential $\Phi(\boldsymbol{d},\boldsymbol{a})$
given a location profile $\boldsymbol{d}$, we have that $\Phi(\bar{\boldsymbol{d}},\boldsymbol{a}_{\bar{\boldsymbol{d}}}^{*})=\max_{\boldsymbol{a}}\Phi(\bar{\boldsymbol{d}},\boldsymbol{a})\geq\Phi(\bar{\boldsymbol{d}},\bar{\boldsymbol{a}})$.
It follows that $\Phi(\bar{\boldsymbol{d}},\boldsymbol{a}_{\bar{\boldsymbol{d}}}^{*})\geq\Phi(\bar{\boldsymbol{d}},\bar{\boldsymbol{a}})>\Phi(\boldsymbol{d}^{*},\boldsymbol{a}_{\boldsymbol{d}^{*}}^{*})$,
which contradicts with that $\Phi(\boldsymbol{d}^{*},\boldsymbol{a}_{\boldsymbol{d}^{*}}^{*})=\max_{\boldsymbol{d}}\Phi(\boldsymbol{d},\boldsymbol{a}_{\boldsymbol{d}}^{*})\geq\Phi(\bar{\boldsymbol{d}},\boldsymbol{a}_{\bar{\boldsymbol{d}}}^{*})$.
Since the joint spatial channel selection and mobility game is a
potential game, we know that the maximum point $\Phi(\boldsymbol{d}^{*},\boldsymbol{a}_{\boldsymbol{d}^{*}}^{*})$ of the potential function must be a Nash equilibrium. \qed

\subsection{Proof of Theorem \ref{thm:For-the-spatialPoA}}\label{proofPoA}
For the ease of exposition, we first define that $E_{n}(\boldsymbol{d})=\max_{m\in\mathcal{M}}\left\{ \log\left(\theta_{m}B_{m,d_{n}}^{n}p_{n}\right)\right\} $
and hence we have $E(\boldsymbol{d})=\min_{n\in\mathcal{N}}\{E_{n}(\boldsymbol{d})\}$.
Since $0<p_{n}<1$, we have $\log(1-p_{n})<0$. It follows from (\ref{eq:payoff2})
that\begin{equation}
U_{n}(\boldsymbol{d},\boldsymbol{a})\leq E_{n}(\boldsymbol{d}).\label{eq:poa1}\end{equation}
Thus,\begin{equation}
\max_{\boldsymbol{a}}\sum_{n\in\mathcal{N}}U_{n}(\boldsymbol{d},\boldsymbol{a})\leq\sum_{n\in\mathcal{N}}E_{n}(\boldsymbol{d}).\label{eq:poa2}\end{equation}
Suppose that $\tilde{\boldsymbol{a}}\in\Xi$ is an arbitrary Nash
equilibrium of the spatial channel selection game. Then at Nash equilibrium,
we must have that\begin{equation}
U_{n}(\boldsymbol{d},\tilde{\boldsymbol{a}})\geq E_{n}(\boldsymbol{d})+\sum_{i\in\mathcal{N}_{n}(\boldsymbol{d})}\log(1-p_{i}).\label{eq:poa3}\end{equation}
Otherwise, the user $n$ always can improve its payoff by choosing
the channel that maximizes $\log\left(\theta_{m}B_{m,d_{n}}^{n}p_{n}\right)$.

According to (\ref{eq:poa2}) and (\ref{eq:poa3}), we then obtain\begin{eqnarray}
\mbox{PoA} & \geq & \frac{\sum_{n\in\mathcal{N}}U_{n}(\boldsymbol{d},\tilde{\boldsymbol{a}})}{\max_{\boldsymbol{a}}\sum_{n\in\mathcal{N}}U_{n}(\boldsymbol{d},\boldsymbol{a})}\nonumber \\
 & \geq & \frac{\sum_{n\in\mathcal{N}}\left(E_{n}(\boldsymbol{d})+\sum_{i\in\mathcal{N}_{n}(\boldsymbol{d})}\log(1-p_{i})\right)}{\sum_{n\in\mathcal{N}}E_{n}(\boldsymbol{d})}\nonumber \\
 & = & 1+\frac{\sum_{n\in\mathcal{N}}\sum_{i\in\mathcal{N}_{n}(\boldsymbol{d})}\log(1-p_{i})}{\sum_{n\in\mathcal{N}}E_{n}(\boldsymbol{d})}\nonumber \\
 & \geq & 1-\frac{\sum_{n\in\mathcal{N}}\sum_{i\in\mathcal{N}_{n}(\boldsymbol{d})}\varpi}{\sum_{n\in\mathcal{N}}E_{n}(\boldsymbol{d})}\nonumber \\
 & \geq & 1-\frac{\sum_{n\in\mathcal{N}}K(\boldsymbol{d})\varpi}{NE(\boldsymbol{d})}\nonumber \\
 & \geq & 1-\frac{K(\boldsymbol{d})\varpi}{E(\boldsymbol{d})}.\label{eq:PoA5}\end{eqnarray}
\qed

\bibliographystyle{ieeetran}
\bibliography{DynamicSpectrum}

\begin{thebibliography}{10}
\providecommand{\url}[1]{#1}
\csname url@samestyle\endcsname
\providecommand{\newblock}{\relax}
\providecommand{\bibinfo}[2]{#2}
\providecommand{\BIBentrySTDinterwordspacing}{\spaceskip=0pt\relax}
\providecommand{\BIBentryALTinterwordstretchfactor}{4}
\providecommand{\BIBentryALTinterwordspacing}{\spaceskip=\fontdimen2\font plus
\BIBentryALTinterwordstretchfactor\fontdimen3\font minus
  \fontdimen4\font\relax}
\providecommand{\BIBforeignlanguage}[2]{{%
\expandafter\ifx\csname l@#1\endcsname\relax
\typeout{** WARNING: IEEEtran.bst: No hyphenation pattern has been}%
\typeout{** loaded for the language `#1'. Using the pattern for}%
\typeout{** the default language instead.}%
\else
\language=\csname l@#1\endcsname
\fi
#2}}
\providecommand{\BIBdecl}{\relax}
\BIBdecl

\bibitem{key-1}
FCC, ``Report of the spectrum efficiency group,'' in \emph{Spectrum Policy Task
  Force}, 2002.

\bibitem{key-2}
I.~Akyildiz, W.~Lee, M.~Vuran, and S.~Mohanty, ``Next generation/dynamic
  spectrum access/cognitive radio wireless networks: a survey,'' \emph{Computer
  Networks}, vol.~50, no.~13, pp. 2127--2159, 2006.

\bibitem{key-3}
B.~Wang, Y.~Wua, and K.~R. Liu, ``Game theory for cognitive radio networks: An
  overview,'' \emph{Computer Networks}, vol.~54, pp. 2537--2561, 2010.

\bibitem{key-21}
N.~Nie and C.~Comaniciu, ``Adaptive channel allocation spectrum etiquette for
  cognitive radio networks,'' in \emph{IEEE Symposium on New Frontiers in
  Dynamic Spectrum Access Networks (DySPAN)}, 2005.

\bibitem{key-4}
D.~Niyato and E.~Hossain, ``Competitive spectrum sharing in cognitive radio
  networks: a dynamic game approach,'' \emph{IEEE Transactions on Wireless
  Communications}, vol.~7, pp. 2651--2660, 2008.

\bibitem{key-22}
``Efficient mac in cognitive radio systems: A game-theoretic approach,''
  \emph{IEEE Transactions on wireless Communications}, vol.~8, pp. 1984--1995,
  2009.

\bibitem{key-16}
L.~M. Law, J.~Huang, M.~Liu, and S.~Y.~R. Li, ``Price of anarchy of cognitive
  {MAC} games,'' in \emph{IEEE Global Communications Conference}, 2009.

\bibitem{key-25}
Z.~Han, C.~Pandana, and K.~J.~R. Liu, ``Distributive opportunistic spectrum
  access for cognitive radio using correlated equilibrium and no-regret
  learning,'' in \emph{IEEE Wireless Communications and Networking Conference
  (WCNC)}, 2007.

\bibitem{key-26}
M.~Maskery, V.~Krishnamurthy, and Q.~Zhao, ``Decentralized dynamic spectrum
  access for cognitive radios: Cooperative design of a non-cooperative game,''
  \emph{IEEE Transactions on Communications}, vol.~57, no.~2, pp. 459--469,
  2009.

\bibitem{key-18}
R.~J. Aumann, ``Correlated equilibrium as an expression of bayesian
  rationality,'' \emph{Econometrica}, vol.~55, pp. 1--18, 1987.

\bibitem{key-27}
A.~Anandkumar, N.~Michael, and A.~Tang, ``Opportunistic spectrum access with
  multiple users: learning under competition,'' in \emph{The IEEE International
  Conference on Computer Communications (Infocom)}, 2010.

\bibitem{key-28}
L.~Lai, H.~Jiang, and H.~V. Poor, ``Medium access in cognitive radio networks:
  A competitive multi-armed bandit framework,'' in \emph{IEEE Asilomar
  Conference on Signals, Systems, and Computers}, 2008.

\bibitem{key-29}
K.~Liu and Q.~Zhao, ``Decentralized multi-armed bandit with multiple
  distributed players,'' in \emph{Information Theory and Applications Workshop
  (ITA)}, 2010.

\bibitem{key-60}
M.~Weiss, M.~Al-Tamaimi, and L.~Cui, ``Dynamic geospatial spectrum modelling:
  taxonomy, options and consequences,'' in \emph{Telecommunications Policy
  Research Conference}, 2010.

\bibitem{key-67}
\BIBentryALTinterwordspacing
C.~Tekin, M.~Liu, R.~Southwell, J.~Huang, and S.~Ahmad, ``Atomic congestion
  games on graphs and their applications in networking,'' to appear in IEEE
  Transactions on Networking. [Online]. Available:
  \url{http://arxiv.org/abs/1011.5384}
\BIBentrySTDinterwordspacing

\bibitem{key-68}
R.~Southwell and J.~Huang, ``convergence dynamics of resource-homogeneous
  congestion game,'' in \emph{proceedings of GameNets Conference}, 2011.

\bibitem{han2008backoff}
S.~Han and N.~Abu-Ghazaleh, ``On backoff in fading wireless channels,''
  \emph{Ad-hoc, Mobile and Wireless Networks}, pp. 251--264, 2008.

\bibitem{kong2004performance}
Z.~Kong, D.~Tsang, B.~Bensaou, and D.~Gao, ``Performance analysis of ieee
  802.11 e contention-based channel access,'' \emph{IEEE Journal on Selected
  Areas in Communications}, vol.~22, no.~10, pp. 2095--2106, 2004.

\bibitem{Zhao2007}
Q.~Zhao, L.~Tong, A.~Swami, and Y.~Chen, ``Decentralized cognitive {MAC} for
  opportunistic spectrum access in ad hoc networks: A pomdp framework,''
  \emph{IEEE Journal on Selected Areas in Communications}, vol.~25, pp.
  589--600, 2007.

\bibitem{kim2008efficient}
H.~Kim and K.~Shin, ``Efficient discovery of spectrum opportunities with
  mac-layer sensing in cognitive radio networks,'' \emph{IEEE Transactions on
  Mobile Computing}, vol.~7, no.~5, pp. 533--545, 2008.

\bibitem{rappaport1996wireless}
T.~Rappaport, \emph{Wireless communications: principles and practice}.\hskip
  1em plus 0.5em minus 0.4em\relax Prentice Hall PTR New Jersey, 1996, vol.~2.

\bibitem{bianchi2000performance}
G.~Bianchi, ``Performance analysis of the ieee 802.11 distributed coordination
  function,'' \emph{IEEE Journal on Selected Areas in Communications}, vol.~18,
  no.~3, pp. 535--547, 2000.

\bibitem{harks2005utility}
T.~Harks, ``Utility proportional fair bandwidth allocation: An optimization
  oriented approach,'' \emph{Quality of Service in Multiservice IP Networks},
  pp. 61--74, 2005.

\bibitem{chen2007contention}
L.~Chen, S.~Low, and J.~Doyle, ``Contention control: A game-theoretic
  approach,'' in \emph{IEEE Conference on Decision and Control}.\hskip 1em plus
  0.5em minus 0.4em\relax IEEE, 2007, pp. 3428--3434.

\bibitem{key-41}
D.~Monderer and L.~S. Shapley, ``Potential games,'' \emph{Games and Economic
  Behavior}, vol.~14, pp. 124--143, 1996.

\bibitem{key-8}
H.~Kushner and G.~Yin, \emph{Stochastic Approximation and Recursive Algorithms
  and Applications}.\hskip 1em plus 0.5em minus 0.4em\relax New York:
  Springer-Verlag,, 2003.

\bibitem{key-90}
L.~Hester and A.~D. Ridley, \emph{The Telecommunications Review}, pp. 44--54,
  2008.

\bibitem{satyanarayanan2001pervasive}
M.~Satyanarayanan, ``Pervasive computing: Vision and challenges,'' \emph{IEEE
  Personal Communications}, vol.~8, no.~4, pp. 10--17, 2001.

\bibitem{balachandran2002hot}
A.~Balachandran, P.~Bahl, and G.~Voelker, ``Hot-spot congestion relief in
  public-area wireless networks,'' in \emph{Proceedings Fourth IEEE Workshop on
  Mobile Computing Systems and Applications}, 2002.

\bibitem{goldenberg2004towards}
D.~Goldenberg, J.~Lin, A.~Morse, B.~Rosen, and Y.~Yang, ``Towards mobility as a
  network control primitive,'' in \emph{Proceedings of the 5th ACM
  international symposium on Mobile ad hoc networking and computing}, 2004.

\bibitem{key-31}
S.~Liew, C.~Kai, H.~Leung, and P.~Wong, ``Back-of-the-envelope computation of
  throughput distributions in csma wireless networks,'' \emph{IEEE Transactions
  on Mobile Computing}, vol.~9, no.~9, pp. 1319--1331, 2010.

\bibitem{key-33}
S.~Zhang, Z.~Shao, and M.~Chen, ``Optimal distributed p2p streaming under node
  degree bounds,'' in \emph{Proceedings of the IEEE International Conference on
  Network Protocols (ICNP)}, 2010.

\bibitem{sevast1957ergodic}
``An ergodic theorem for markov processes and its application to telephone
  systems with refusals,'' \emph{Theory of Probability and its Applications},
  vol.~2, p. 104, 1957.

\bibitem{chandy1977product}
K.~Chandy, J.~Howard~Jr, and D.~Towsley, ``Product form and local balance in
  queueing networks,'' \emph{Journal of the ACM (JACM)}, vol.~24, no.~2, pp.
  250--263, 1977.

\bibitem{vazquez2006modeling}
A.~V{\'a}zquez, J.~Oliveira, Z.~Dezs{\"o}, K.~Goh, I.~Kondor, and
  A.~Barab{\'a}si, ``Modeling bursts and heavy tails in human dynamics,''
  \emph{Physical Review E}, vol.~73, no.~3, p. 036127, 2006.

\bibitem{key-30}
S.~Boyd and L.~Vandenberghe, \emph{Convex optimization}.\hskip 1em plus 0.5em
  minus 0.4em\relax Cambridge university press, 2004.

\bibitem{key-99}
R.~S. Sastry, V.~V. Phansalkar, and M.~A.~L. Thathachar, ``Decentralized
  learning of nash equilibria in multi-person stochastic games with incomplete
  information,'' \emph{IEEE Transactions on Systems Science and Cybernetics},
  vol.~2, pp. 769 -- 777, 1994.

\bibitem{schoen1994lectures}
R.~Schoen and S.~Yau, \emph{Lectures on differential geometry}.\hskip 1em plus
  0.5em minus 0.4em\relax International press Boston, 1994, vol.~2.

\bibitem{kleinberg2009multiplicative}
R.~Kleinberg, G.~Piliouras, and {\'E}.~Tardos, ``Multiplicative updates
  outperform generic no-regret learning in congestion games,'' in
  \emph{Proceedings of the 41st annual ACM symposium on Theory of computing},
  2009.

\end{thebibliography}


%

\end{document}